\documentstyle[preprint,aps,eqsecnum,epsf]{revtex}

\begin{document}
\setcounter{page}{1}

\draft

\title{
 Pseudoscalar pole terms in the
 hadronic light-by-light  \\
  scattering contribution 
  to muon $ g - 2 $  }

\author{
 M. Hayakawa$ ^1$ \footnotemark[1] and
 T. Kinoshita$ ^{2}$ \footnotemark[2] 
 }
\footnotetext[1]
{
 Electronic address: hayakawa@theory.kek.jp
}
\footnotetext[2]
{
 Electronic address: tk@hepth.cornell.edu
}

\address{
 $ ^1 $  Division of Theoretical Physics, KEK, \\
         Tsukuba, Ibaraki, 305 Japan \\
 $ ^2 $  Newman Laboratory, Cornell University, \\
         Ithaca, New York, 14853 USA }

\preprint{\begin{minipage}{4cm}
           KEK-TH-530 \\
           hep-ph/9708227 \\
           August 1997 \\
          \end{minipage}
         }

\date{September 23, 1997}

\maketitle

\begin{abstract}
 The pseudoscalar pole contribution is 
the dominant source of the $ {\cal O}(\alpha^3) $
hadronic light-by-light scattering effect in muon $g - 2$.
 We have examined this contribution
taking account of the off-shell structure
of the pseudoscalar-photon-photon anomaly vertex
deduced from available experimental data.
 Our work leads to an improved estimate, 
$ -79.2\ (15.4) \times 10^{-11}$, 
for the total hadronic
light-by-light scattering contribution 
to the muon $g - 2$.\\
\end{abstract}


\narrowtext
%

\section{Introduction}
\label{sec:intro}

  New measurement of the anomalous magnetic moment
of the muon (muon anomaly)
$a_\mu = \frac{1}{2} (g_\mu - 2)$
is underway at Brookhaven National Laboratory
\cite{miller}.
  The anticipated level of precision,
$40 \times 10^{-11}$,
is more than 20 times higher
than that of the best previous result
\cite{CERN}
\begin{equation}
 a_\mu ({\rm exp})
  = 1\ 165\ 923\ (8.5) \times 10^{-9} ,
\end{equation}
where the numerals in parentheses represent
the uncertainties
in the last digits of the measured value.
 The primary purpose
of the new muon $g - 2$ experiment
is to verify the presence
of the electroweak contribution.
 Other effects of potential interest are those of 
supersymmetric particles
\cite{SUSY,SUSYdata}
and leptoquarks \cite{leptoquark,lqdata}.

 At present
the standard model prediction of $a_\mu$ is

\begin{equation}
 a_\mu ({\rm th})
  = 116\ 591\ 714\ (96) \times 10^{-11}.
 \label{eq:prediction}
\end{equation}
 This consists of five parts: 

\noindent
(i) Pure QED contribution
\footnote{
 Eq. (\ref{amuQED})
is obtained from the measured value of
the electron anomaly $a_e$ \cite{vandyck}
$minus$ a small correction to $a_e$
due to muon, hadron, and weak interactions
\cite{alpha}, $plus$ the terms of $a_\mu$ dependent
on the electron and tau masses
evaluated using the fine structure constant
obtained from the electron anomaly \cite{alpha}. }
\begin{equation}
 a_\mu ({\rm QED})
  = 116\ 584\ 705.7\ (1.9) \times 10^{-11}.
 \label{amuQED}
\end{equation}

\noindent
(ii) Hadronic vacuum polarization contribution 
obtained mainly
from the measured hadron production cross section
in $e^+ e^-$ collisions 
\cite{K-had,Gourdin,groups,worstell}.
 We quote here the latest evaluation
that includes additional information
obtained from the analysis
of hadronic tau decay data \cite{Alemany,CLEO-tau}:
\begin{equation}
 a_\mu ({\rm had.v.p.})
  = 7\ 011\ (94) \times 10^{-11}.
 \label{had.v.p. contr}
\end{equation}

\noindent
(iii) Higher order
hadronic vacuum polarization effect
\cite{krause}:

\begin{equation}
 a_\mu ({\rm higher~had.v.p.})
  = - 101\ (6) \times 10^{-11}.
 \label{higherhad.v.p. contr}
\end{equation}

\noindent
(iv) Hadronic light-by-light scattering contribution
\cite{HKS}:
\begin{equation}
 \displaystyle{
    a_\mu(\mbox{had. l-l})
  }
  =
  \displaystyle{
   -52\ (18) \times 10^{-11} 
  },
 \label{HKSsum}
\end{equation}
and a similar result
obtained independently in Ref. \cite{Bijnens}.

\noindent
(v) Electroweak contribution of
one \cite{weak} and two \cite{weak2} loop orders:
\begin{equation}
 a_\mu({\rm weak})=151\ (4) \times 10^{-11}.
\label{weakeffect}
\end{equation}

 The current uncertainty in the theoretical value of $a_\mu$ 
comes mostly from the hadronic contribution.
 It must be improved by at least a factor of two
before we can extract useful physical information
from the new high precision measurement
and impose strong constraints on
various candidates
for possible extension of the standard model.

 The hadronic contribution appears for the first time
in the order $\alpha^2 $
as the effect of hadronic vacuum polarization.
(See Fig. 1 of Ref. \cite{K-had} for the Feynman graphs 
responsible for such a contribution.)
 Fortunately, the contribution of this type does not require
explicit $ab~initio$ calculation based on QCD,
since it is precisely calculable from the measured hadron
production cross section in $e^+e^-$ collisions
\cite{K-had,Gourdin,groups}.
 Future measurements
at VEPP-2M, VEPP-4M, DA$\Phi$NE and BEPS,
as well as analysis of
the hadronic tau decay data 
are expected to reduce the uncertainty of this contribution
to the level that satisfies our need
\cite{worstell,Franzini}.

 The contribution
of the hadronic light-by-light scattering subdiagram
is much smaller but 
is potentially a source of more serious problem because 
it has been difficult to express it in terms of 
experimentally accessible observables.
 At present
it depends entirely on theoretical consideration.
 This contribution
 has been estimated recently by two groups 
within the framework of
chiral perturbation theory and the $1/N_c$ expansion
\cite{HKS,Bijnens}.
 The leading terms arise from three types of diagrams
shown in Fig. \ref{fig:diagram}:
(a) pion-loop contribution,
(b) pseudoscalar pole contribution,
and (c) quark-loop contribution.
 The results obtained in Ref. \cite{HKS}
for these diagrams are
 \begin{eqnarray}
   \left.
    a_\mu (a) 
   \right|_{\rm HKS}
&=& \ \,-4.5\ (8.1)\ \ \times 10^{-11},   \nonumber   \\
   \left.
    a_\mu (b) 
   \right|_{\rm HKS}
&=& -57.5\ (11.4) \times 10^{-11},   \nonumber   \\
   \left.
    a_\mu (c) 
   \right|_{\rm HKS}
&=& \ \ \ \, 9.7\ (11.1) \times 10^{-11}.
 \label{HKSabc}
\end{eqnarray}
 They add up to (\ref{HKSsum}).
 A small axial-vector contribution 
\begin{equation}
 \displaystyle{
   \left.
    a_\mu(\mbox{axial-vector})
   \right|_{\rm HKS}
  }
  =
  \displaystyle{
   -1.74 \times 10^{-11} 
  },
 \label{HKSaxial}
\end{equation}
was also obtained 
but not included in (\ref{HKSsum}) .
( See eq. (4.32) of Ref. \cite{HKS}.)
 The corresponding results obtained in \cite{Bijnens} were
 \begin{eqnarray}
   \left.
    a_\mu (a) 
   \right|_{\rm BPP}
&=& -19\ (13) \times 10^{-11},   \nonumber   \\
   \left.
    a_\mu (b) 
   \right|_{\rm BPP}
&=& -85\ (13) \times 10^{-11},   \nonumber   \\
   \left.
    a_\mu (c) 
   \right|_{\rm BPP}
&=& \ \,\,\,21\ (3) \ \,\times 10^{-11}.
 \label{BPPabc}
\end{eqnarray}
 The effects of axial-vector
and scalar poles were also considered
in Ref. \cite{Bijnens}:
 \begin{eqnarray}
   \left.
    a_\mu (\mbox{axial-vector}) 
   \right|_{\rm BPP}
&=& -2.5\ (1.0) \times 10^{-11},   \nonumber   \\
   \left.
    a_\mu ({\rm scalar}) 
   \right|_{\rm BPP}
&=& -6.8\ (2.0) \times 10^{-11}.   
 \label{BPPaxsc}
\end{eqnarray}
 Summing up (\ref{BPPabc}) and (\ref{BPPaxsc}),
they obtained
\begin{equation}
 \displaystyle{
   \left.
    a_\mu(\mbox{had. l-l})
   \right|_{\rm BPP}
  }
  =
  \displaystyle{
   -92\ (32) \times 10^{-11}
  }. \label{BPPsum}
\end{equation}

 The reasons for the difference between (\ref{HKSabc}) and
(\ref{BPPabc}) are as follows:
 For the pion-loop contribution
$a_\mu (a)$
it is due to the fact that the effective Lagrangian
responsible for (\ref{BPPabc}) has
a non-derivative $\rho^0 \rho^0 \pi^+ \pi^-$ coupling
while the corresponding term is absent in (\ref{HKSabc}),
which is obtained from a vector meson dominance (VMD) model
with hidden local symmetry \cite{bando}.
 The absence of this term in the latter
is a direct consequence of the fact that 
it satisfies the Ward-Takahashi identity \cite{HKS} 
and a soft pion theorem
for $V^0 \pi$ scattering amplitude \cite{hayakawa}.

 In Ref. \cite{Bijnens} an effective chiral Lagrangian
was proposed which reproduces
the earlier VMD result of Ref. \cite{K-had} in terms of
an interaction term consisting of an
infinite series of derivatives
of the pseudoscalar meson.
 These higher derivative terms are accompanied
by a mass scale $M_C$,
which, after their resummation,
becomes just a pole mass
corresponding to a vectorial degree of freedom.
 Their result will be justified
if that $M_C$ can be
identified with the vector meson mass $M_\rho$.
 However, the analysis of the low energy behavior
of the $\rho^0 \pi^+ $ scattering
prevents us from interpreting the above degree of freedom
as being associated
with the physical vector meson \cite{hayakawa}.
 The scale $M_C$ will have to be associated
with another physical degree of freedom,
which is heavier than $M_\rho$.
 For such an $M_C$
the magnitude of the charged pion loop contribution
based on their model
will become smaller than that given in \cite{Bijnens}.
In this model the relation between the scale $M_C$
and the resulting value
of the charged pion loop contribution
$\left. a_\mu(a)\right|_{\rm BPP}$
in (\ref{BPPabc}) is not transparent.
 For this reason we will henceforth choose
$\left. a_\mu(a)\right|_{\rm HKS}$ of (\ref{HKSabc})
as the contribution of Fig. \ref{fig:diagram} (a).

 The difference in the evaluation of $a_\mu (b)$
is mainly due to the fact that
only the contributions of the $\pi^0$ pole and $\eta$ pole
are taken into account in (\ref{HKSabc}) 
whereas (\ref{BPPabc}) includes also the contribution
of $\eta^\prime$,
which turned out to be  not negligible \cite{Bijnens}.
 When the $\eta^\prime$ contribution
is added to (\ref{HKSabc}) (see Sec. 
\ref{sec:formfactor}),
the remaining difference for $a_\mu (b)$ is no longer large
and reflects mainly the ambiguity and difficulty in
carrying out the chiral perturbation theory estimate 
beyond the momentum range of several hundred MeV.

 Similarly, the difference between $a_\mu (c)$
of (\ref{HKSabc}) and (\ref{BPPabc})
originates from the difficulty in estimating the contribution
from large momentum region.

 As is seen from (\ref{HKSabc}) and ({\ref{BPPabc}),
the most important contribution 
comes from the diagrams of type (b) in which 
neutral pseudoscalar mesons $P$ 
( $ P = \pi^0$, $\eta$ and $\eta^\prime$ )
propagate between two
$ P \gamma \gamma $ vertices,
as shown
in Fig. \ref{fig:pseudoscalarpole}.
 As is well known, 
the chiral anomaly for $\pi_0$, for instance,
can be expressed by the effective interaction
\begin{equation}
{\cal L} = -{\alpha \over {8\pi f_\pi}} \pi^0
          \epsilon^{\mu \nu\lambda \sigma}
          F_{\mu \nu} F_{\lambda \sigma} ,
\label{effectiveL}
\end{equation}
where $f_\pi \simeq 93 $ MeV is the pion decay constant,
in the lowest order of chiral expansion.
 When applied to the calculation
of $a_\mu (b)$, however, this Lagrangian 
leads to an ultraviolet-divergent result.
 This divergence
arises from the triangular photon-photon-muon loop 
which is obtained
by reducing the $\pi^0 \gamma \gamma$ vertex, 
represented by a large shaded blob on the left-hand side
of Fig. \ref{fig:pseudoscalarpole}(a), to a point.
(Fig. \ref{fig:pseudoscalarpole}(b)
is convergent in the same limit.)
 It is a signal that the local interaction
(\ref{effectiveL})
is not applicable to photons and pions far off mass shell. 
 In fact, such a triangular diagram
will have a damping behavior in the underlying QCD theory,
which protects its contribution to the muon $g-2$
from diverging.

 In Ref. \cite{HKS}, 
four models were considered to examine the effect of
various assumptions on the off-shell behavior of the
$\pi^0 \gamma \gamma$ form factor:

\noindent
~~~($b_1$) vector meson dominance (VMD) model, 
\footnote{The VMD model can be justified
within the hidden local symmetry
picture of the chiral model \cite{bando}.}

\noindent
~~~($b_2$) quark triangular loop (QTL) model,

\noindent
~~~($b_3$) QTL model combined with the VMD model,

\noindent
~~~($b_4$) extended Nambu-Jona-Lasinio (ENJL) model
           \cite{ENJL}.

\noindent
 The contribution 
$a_\mu (b; \pi^0 )$ 
of Fig. \ref{fig:pseudoscalarpole} 
evaluated by these models,
and given by Eqs. (4.1), (4.3) and (4.4),
and Table V (with $g_A = 0.5$, $M_C = 1$ GeV)
of Ref. \cite {HKS}, are reproduced here:
 \begin{eqnarray}
   a_\mu (b_1 ; \pi^0 ) &=& - 55.60 \times 10^{-11} ,
     \nonumber   \\
   a_\mu (b_2 ; \pi^0 ) &=& - 86.90 \times 10^{-11} ,
     \nonumber   \\
   a_\mu (b_3 ; \pi^0 ) &=& - 33.76 \times 10^{-11} ,
     \nonumber   \\
   a_\mu (b_4 ; \pi^0 ) &=& - 42.84 \times 10^{-11} .  
\label{pi0models}
\end{eqnarray}
 Similar calculations were carried out 
in Ref. \cite{HKS} 
for the $\eta$ resonance, too.
 The final value of $a_\mu (b)$ reported in Ref. \cite{HKS}
and quoted in (\ref{HKSabc}) is based mainly
on the model $(b_4)$.

 After completion of these calculations 
an experimental measurement of the $P \gamma \gamma^*$
form factor,
where $P$ stands for $\pi^0 ,~ \eta $ or $ \eta'$, 
came to our attention \cite{CELLO,CLEO,CLEO-2}.
 This is very important in the sense that it 
opens up the possibility of evaluating $a_\mu (b)$
utilizing the experimental information instead of relying  
solely on theoretical consideration.
 This may enable us to reduce significantly the
uncertainty in the evaluation of $a_\mu (b)$.

 The purpose of this paper is to amplify the preliminary
discussion in Ref. \cite{HKS} about the implication of
the measurements
and discuss in full 
the effect of the measured $P \gamma \gamma^*$ form
factor (Sec. \ref{sec:comparison})
and possible impacts of the yet-to-be-measured
$P \gamma^* \gamma^*$ form factor
on the evaluation of $a_\mu (b)$
(Sec. \ref{sec:formfactor}).
 Before going into these sections,
we review the theoretical aspects of the asymptotic behavior
of the form factor of our interest
in Sec. \ref{sec:asymptotic}.
 The last section (Sec. \ref{sec:discussion})
will be devoted
to the summary and discussion of our results. 

\section{Theoretical off-shell
         $\pi^0 \gamma^* \gamma^* $ form factor}
\label{sec:asymptotic}

 Let us write 
the invariant $P \gamma^* \gamma^*$ amplitude as
\begin{eqnarray}
 &&
 \displaystyle{
  {\cal M}(\gamma^*(p_1,\lambda_1) \gamma^*(p_2,\lambda_2)
           \rightarrow P(q))
 } \nonumber \\
 && \quad \quad
 \displaystyle{
  = \epsilon^\mu_{\lambda_1}(p_1)
    \epsilon^\nu_{\lambda_2}(p_2)\,
    \varepsilon_{\mu\nu\alpha\beta}\,p_1^\alpha p_2^\beta\,
    M_P(p_1^2,\, p_2^2,\, q^2 = m_P^2)
 },
 \label{eq:inv-amp}
\end{eqnarray}
and define the form factor $F_P (p_1^2,\, p_2^2,\, q^2)$
by
\begin{eqnarray}
 F_P(p_1^2,\, p_2^2,\, q^2) &=&
  \displaystyle{
    \frac{\pi f_P}{\alpha}
    M_P (p_1^2,\, p_2^2,\, q^2)
  }.
\end{eqnarray}
In the chiral limit ($q^2 = m_P^2 = 0$),
this is normalized as

\begin{equation}
 F_P (0,\, 0,\, 0) = 1.
  \label{eq:normalization}
\end{equation}

 Let us first examine the off-shell behavior
of the $\pi^0 \gamma^* \gamma^* $ form factor
in the quark triangular loop (QTL) model.
 In this model, we find (ignoring isospin violation)
\begin{eqnarray}
 F_{\pi^0}^{\rm QTL} (p_1^2,\, p_2^2,\, q^2) &=&
 \displaystyle{
  I_{m_u^2}(p_1^2,\, p_2^2,\, q^2)
 } \nonumber \\
 &\equiv&
 \displaystyle{
  \int [dz]
   {2 m_u^2 \over
   {m_u^2 - z_2 z_3 p_1^2 - z_3 z_1 p_2^2 - z_1 z_2 q^2}} 
 },
\label{functionF0}
\end{eqnarray}
where $[dz] = dz_1 dz_2 dz_3 \delta (1 - z_1 - z_2 - z_3 )$,
and $m_u \sim 300$ MeV is the constituent mass
of up (and down) quark \cite{Manohar}.
 For $p_1^2 = p_2^2 = q^2 = 0$
this function reduces to 
(\ref{eq:normalization}).
 Carrying out the integration one can readily find that
\begin{eqnarray}
 F_{\pi^0}^{\rm QTL}(p_1^2,\, p_2^2,\, 0)
  &=&
  \displaystyle{
   - {m^2 \over {p_1^2 - p_2^2 }}
   \left[ 
    \left\{
     \ln
      \left( 
       { \sqrt{4 m_u^2 - p_1^2 } + \sqrt{-p_1^2 } }
        \over
       { \sqrt{4 m_u^2 - p_1^2 } - \sqrt{-p_1^2 } }
      \right)
    \right\}^2
   \right.
  } \nonumber \\
  && \quad \quad \quad \quad
  \displaystyle{
   \left.
   -   
    \left\{
     \ln
      \left( 
       { \sqrt{4 m_u^2 - p_2^2 } + \sqrt{-p_2^2 } }
        \over
       { \sqrt{4 m_u^2 - p_2^2 } - \sqrt{-p_2^2 } }
      \right)
    \right\}^2\,\,
   \right] .
  }
  \label{analytic}
\end{eqnarray}
 For large $p_1^2 $ with $p_2^2 = 0$ and $q^2 = 0$
this has the asymptotic behavior of the form
\begin{equation}
 F_{\pi^0}^{\rm QTL} (p_1^2,\, 0,\, 0)
  \sim
   {m_u^2 \over {-p_1^2}}
     \left\{
       \ln \left( \frac{-p_1^2}{m_u^2} \right)
     \right\}^2 . 
 \label{case1}
\end{equation}
 For large $p_1^2 \sim p_2^2 $ and $q^2 = 0$ we find
\begin{equation}
 F_{\pi^0}^{\rm QTL} (p_1^2,\, p_1^2,\, 0)
  \sim
    {2m_u^2 \over {-p_1^2}}
      \ln \left( \frac{-p_1^2}{m_u^2} \right).
 \label{case2}
\end{equation}
 Note that both (\ref{case1}) and (\ref{case2})
have the same power behavior at large momentum transfer,
differing only in the logarithmic factor. 

 Let us define the function ${\cal F}_{\pi^0}$ by
\begin{eqnarray}
 {\cal F}_{\pi^0}(Q^2) &=&
 \displaystyle{
  \frac{1}{4\pi\alpha}
  \left|
    M_{\pi^0}(-Q^2,\,0,\,0)
  \right|
 } \nonumber \\
 &=&
 \displaystyle{
  \frac{1}{4 \pi^2 f_\pi}
   \left| F_{\pi^0}(-Q^2,\,0,\,0) \right|
 }\ .
 \label{curlyF}
\end{eqnarray}
 Fig. \ref{fig:formfactor} shows 
the momentum dependence of
$ Q^2 {\cal F}_{\pi^0}(Q^2) $
for the cases ($b_1$), ($b_2$) and ($b_3$) 
together with the experimental data reported
by CLEO collaboration \cite{CLEO-2}.
 This shows
that the VMD model ($b_1$) fits the experimental data
particularly well.
 The model ($b_2$) gives slower damping for higher momenta.
 This is because of the extra $(\ln (-p_1^2))^2$ factor 
found in (\ref{case1}) which compensates
for the leading damping $1/p_1^2$ to some extent.
 The model ($b_3$), on the other hand,
predicts rapid decrease
due to the stronger damping factor
$(\ln (-p_1^2))^2/p_1^4$.

 Nonperturbative analysis of
 asymptotic behavior of the exact form factor 
$F_{\pi^0}(p_1^2,\, p_2^2,\, q^2 )$ 
has also been carried out by several 
methods \cite{Manohar,Brodsky,Gerard}.
 Various corrections to the result of \cite{Brodsky}
have also been considered:
Ref. \cite{Cao} discusses
the effect of the parton transverse momentum.
 Ref. \cite{Jacob} considers the
gluonic radiative corrections to \cite{Cao}. 
 Ref. \cite{Radyushkin} derived
an asymptotic formula
applying the QCD sum rule \cite{Shifman}.
 These corrections have been compared with the data in
\cite{CLEO-2}.
 Here, we concentrate on the methods
of Refs. \cite{Manohar,Brodsky,Gerard}
since they are more directly related to the
consideration of this paper.

 For $p_2^2 = q^2 = 0$ all three methods 
\cite{Manohar,Brodsky,Gerard}
predict
the same leading power of momentum for large $p_1^2$.
 But the coefficients found are different:
\begin{eqnarray}
 \displaystyle{
   \lim_{-p_1^2 \rightarrow \infty}
   F_{\pi^0}(p_1^2 , 0, 0)
 }
 &=&
 \displaystyle{
  \frac{3}{N_c} \frac{2\pi^2 f_\pi^2}{-p_1^2},
 }~~~~~~~~~~~~~~~~~~ \cite{Brodsky}  \label{Brodsky}  \\ 
 &=&
 \displaystyle{
  \frac{2}{N_c} \frac{2\pi^2 f_\pi^2}{-p_1^2},
 }~~~~~~~~~~~~~~~~~~ \cite{Manohar}  \label{Manohar}  \\
 &=&
 \displaystyle{
  \frac{1}{N_c} \frac{2\pi^2 f_\pi^2}{-p_1^2}.
 }~~~~~~~~~~~~~~~~~~ \cite{Gerard}   \label{Gerard}
\end{eqnarray}
 The asymptotic behavior of 
$F_{\pi^0}(p_1^2,\, p_2^2,\, 0)$ 
for large $p_1^2 = p_2^2$ has also been studied
nonperturbatively \cite{Novikov}:
\begin{equation}
  \lim_{-p_1^2 \rightarrow \infty}
    F_{\pi^0}(p_1^2,\, p_1^2,\, 0) =
  \frac{1}{N_c} \frac{2\pi^2 f_\pi^2}{-p_1^2} .
\label{novikov}
\end{equation}
 Ref. \cite{Gerard} states in addition that (\ref{Gerard}) 
holds for all values of $p_2^2$ including
$p_2^2 \sim p_1^2$.

 Different asymptotic behaviors of (\ref{Brodsky}),
(\ref{Manohar}) and (\ref{Gerard})
presumably reflect different physical assumptions.
 Eq. (\ref{Brodsky}) is obtained by appealing
to the parton picture in the infinite momentum frame.
 The leading momentum power
dependence there
comes from the scaling behavior of the pion wave function.
 However the coefficient depends
on the long distance aspect of the pion,
which is vulnerable to theoretical prejudices.
 According to literatures \cite{CLEO-2,Cao,Jacob}
the ansatz adopted by Brodsky and Lepage
for the pion wave function
is not badly in conflict with the experiment.

 Eq. (\ref{Manohar}) was obtained
using the operator product expansion (OPE) technique,
within the approximation
of the dominance of operators of lowest dimension and twist
\cite{Manohar}.
  The reliability of OPE
and truncation of operators
depend on the detailed prescription
of how the large momentum limit is taken.
 In its application to the current issue
the form factor is expanded in
$Q^2 = -(p_1 + p_2)^2 /4$
and $\omega = (p_1^2 - p_2^2)/Q^2$.
 The leading power and its coefficient
can be determined including long distance effect
only when the absolute magnitude of $\omega$ is small.
 This is the case for large $p_1^2 \sim p_2^2$
but not for large $p_1^2$ with $p_2^2$ fixed to 0.
 In the latter case the estimate based on the OPE is not trustworthy
since $\omega$ will not be restricted to small values
and the long distance contributions associated
with higher power of $\omega$ might become
as significant as the leading term
\cite{Manohar}.

 The derivation of (\ref{Gerard}) relies
on the Bjorken-Johnson-Low theorem \cite{Bjorken}.
 This theorem is usually used
to {\it define} the commutator of operators
by postulating an appropriate
asymptotic behavior for some correlation function.
 The asymptotic behavior 
in (\ref{Gerard}), on the other hand, is derived from
the Bjorken-Johnson-Low (BJL) limit 
of the matrix element
of the commutator of the electromagnetic currents
\cite{Gerard}.
 The use of equal-time commutation relation
of the quark fields
turns this commutator into the axial current.
 This enabled them to express
the relevant matrix element in terms of
the pion decay constant $ f_\pi $.
However, this approach may not be reliable because
the canonical equal-time commutation relation
does not necessarily 
lead to the correct anomaly relation \cite{Christos}.
 Whether this caution is also relevant for the evaluation of
the commutator of the electromagnetic currents or not
will not be pursued here.
 Irrespective of the validity of
derivation of (\ref{Gerard})
in \cite{Gerard},
we note
that the asymptotic behavior of $\pi^0 \gamma \gamma^*$
in the time-like region will differ significantly
from that measured in the space-like region.
  At present we cannot find reason
to support the possibility
that such a non-trivial continuation
occurs between the two regions.

 In order to compare the theoretical predictions of 
the $\pi^0 \gamma \gamma^*$ form factor with experiment,
it is necessary to interpolate them
from the asymptotic region to
$p_1^2 = 0$, where they are fixed by the anomaly condition.
 Ref. \cite{Brodsky} proposed a one-parameter
interpolation formula
\begin{equation}
 {\cal F}_{\pi^0}(Q^2) 
    = {1 \over {4 \pi^2 f_\pi}}
      {1 \over {1 + Q^2/(8 \pi^2 f_\pi^2 )}} ,
\label{BLinterpolation}
\end{equation}
where $Q^2 = -p_1^2$.
 The formula (\ref{BLinterpolation})
has the same $Q^2$ dependence as
(\ref{experimentalpiformfactor})
if we choose $8\pi^2 f_\pi^2 = M^2$.
Thus this interpolation reproduces
the $Q^2$ dependence very well.
 However, this also fixes the value of ${\cal F}_{\pi^0}(0)$
which differs
from the observed value by a factor of $\sim$ 1.2.
 In this sense (\ref{BLinterpolation})
does not fully reproduce the experimental data.
 Interpolations of (\ref{Manohar}) and (\ref{Gerard})
will be even more difficult in this respect.

\section{ $P \gamma\gamma^*$ form factor}
\label{sec:comparison}
\subsection{Experimental results}
\label{subsec:comparison}

 Depicted in Fig. \ref{fig:asy} are the function
$Q^2 {\cal F}_{\pi^0}(Q^2)$,
corresponding to the asymptotic form factors
(\ref{Brodsky}),
(\ref{Manohar}) and (\ref{Gerard}),
and the interpolation function (\ref{BLinterpolation}),
together with
the observed $\pi^0 \gamma \gamma^*$ form factor
provided by the recent CLEO data \cite{CLEO-2}.
 It is seen that
none of the asymptotic behaviors shown there
is fully consistent with the new data provided by CLEO
for $f_\pi = 93$ MeV.

 On the other hand, it was noted \cite{CLEO} that
the experimental data can be represented
very well by the empirical formula
\begin{equation}
 {\cal F}_{\pi^0} (Q^2)
  = 
  \sqrt{
   {{64\pi\,\Gamma ( \pi^0 \rightarrow \gamma \gamma)} 
    \over {(4\pi \alpha )^2 m_{\pi^0}^3}}
  } {1 \over {1 + (Q^2/\Lambda_{\pi^0}^2 )}},
 \label{experimentalpiformfactor}
\end{equation}
if one chooses $\Lambda_{\pi^0} \simeq 0.77$ GeV  
and $\Gamma (\pi^0 \rightarrow \gamma \gamma )$
= 7.78 (56) eV.
 The uncertainty in $\Gamma$ is about $\pm 7 \%$ \cite{PDG}.
 Note that this $\Lambda_{\pi^0}$ is nearly identical
with the physical $\rho $ mass.

 Similarly,
the experimental data for the $\eta$ and $\eta^\prime$ 
form factors can be represented well by 
\begin{equation}
 {\cal F}_{P} (Q^2) =
   \sqrt{
    {{64\pi\,\Gamma (P \rightarrow \gamma \gamma)} 
     \over {(4\pi \alpha )^2 m_P^3}}
   } 
   {1 \over {1 + (Q^2/\Lambda_P^2 )}}, 
 \label{experimentaletaformfactor}
\end{equation}
for $P = \eta\ {\rm and}\ \eta^\prime$,
with the choice of $\Lambda_\eta \simeq 0.77$ GeV
and $\Lambda_{\eta^\prime} \simeq 0.85$ GeV.
Here, to take account of non-zero mass of meson properly,
we have modified (\ref{curlyF}) as
\begin{eqnarray}
 {\cal F}_P(Q^2) &=&
 \displaystyle{
  \frac{1}{4\pi\alpha}
  \left|
    M_P(-Q^2,\,0,\,m_P^2)
  \right|
 }.
 \label{eq:cleo-f}
\end{eqnarray}
The life-times chosen are 
$\Gamma (\eta \rightarrow \gamma \gamma)
 = 0.46\ (4)\,{\rm keV}$, and 
$\Gamma (\eta^\prime \rightarrow \gamma \gamma)
 = 4.26\ (19)\,{\rm keV}$. \cite{PDG}
 
\subsection{Effect of the $P\gamma\gamma^*$ form factor
            on muon $g - 2$}
\label{subsec:effectofpgg}

 Let us now evaluate the pseudoscalar pole contribution
to the muon $g - 2$,
assuming that all virtual photons
are modified by the $P \gamma^* \gamma^*$ form factor
in accord with the VMD model.
 We consider two models.
 The first model (${\rm VMD}_A$) assumes
that the vector meson dominance is
realized
in terms of the $\rho$ and $\phi$ mesons of the form
\begin{eqnarray}
 M_P(p_1^2,\, p_2^2,\, q^2) &=&
 \displaystyle{
  \frac{1}{f_P} \frac{\alpha}{\pi}
  \left\{
   F_\rho^{\rm VMD} (p_1^2,\, p_2^2,\, q^2)
   + \kappa_\phi(P) F_\phi^{\rm VMD} (p_1^2,\, p_2^2,\, q^2)
  \right\}
 },
 \label{VMD1}
\end{eqnarray}
with
\begin{equation}
 F_V^{\rm VMD} (p_1^2,\, p_2^2,\, q^2)
  = \frac{M_V^2}{M_V^2 - p_1^2}
    \frac{M_V^2}{M_V^2 - p_2^2},
 \label{eq:VMD-form}
\end{equation}
for $ V = \rho$ and $\phi$,
while $\kappa_\phi(P)$
is treated as an adjustable parameter.
 Here we have adopted the approximation $M_\rho \simeq M_\omega$.
 Note that $f_P$ appearing in (\ref{VMD1}) is
not the usual decay constant.
 Rather it should be regarded
as an ``effective'' decay constant.
 When we fit (\ref{VMD1}) 
with the experimental data 
we use the expression
\begin{eqnarray}
 {\cal F}_{\pi^0}(Q^2) &=& 
  \displaystyle{
    \sqrt{
     \frac{64\pi\,\Gamma(P \rightarrow \gamma \gamma)}
          {(4\pi\alpha)^2 m_P^3}
    }
    \times
    \frac{ \left| M_P(-Q^2,\,0,\,m_P^2) \right|}
         { \left| M_P(0,\,0,\,m_P^2) \right|}
  },
 \label{eq:f-m}
\end{eqnarray}
which follows from (\ref{eq:cleo-f}) and 
\begin{eqnarray}
 \Gamma(P \rightarrow \gamma\gamma) &=&
  \displaystyle{
    \frac{m_P^3}{64 \pi}\,
    \left| M_P(0,\,0,\,m_P^2) \right|^2
  }.
\end{eqnarray}
 Note that fitting of (\ref{VMD1}) with data does not involve $f_P$.
 The value of $f_P$ itself
will be determined
once we fix $\kappa_\phi(P)$ from the experiment
for a given $\Gamma(P \rightarrow \gamma\gamma)$
through the relation
\begin{eqnarray}
 \Gamma(P\rightarrow \gamma \gamma) &=&
 \displaystyle{
  \frac{m_P^3}{64\pi}
  \left|
    M_P(0,0,m_P^2)
  \right|^2
 }  \nonumber \\
 &=&
 \displaystyle{
  \frac{1}{f_P^2}\,
  \frac{m_P^3}{64\pi}\,
  \left(
   \frac{\alpha}{\pi}
  \right)^2
  \left|
    F_\rho^{\rm VMD} (0,0,m_P^2)
    + \kappa_\phi(P)
    F_\phi^{\rm VMD} (0,0,m_P^2)
  \right|^2
 }.
 \label{eq:gamma-f}
\end{eqnarray}

 The second model (${\rm VMD}_B$) assumes
that the vector meson dominance is
realized in terms of the experimental pole mass
$\Lambda_P$ obtained in \cite{CLEO}.
 We vary $\Lambda_P$ within its error bars.

 In the model ${\rm VMD}_A$ (or 
${\rm VMD}_B$)
the parameter $\kappa_\phi(P)$ (or $\Lambda_P$) is determined
by fitting to the experimental data \cite{CLEO-2}
for each pseudoscalar meson $P$
and a fixed value of $\Gamma(P \rightarrow \gamma\gamma)$
using the CERN library routine MINUIT.
 Here the confidence level of $1\sigma$ is imposed
to extract the uncertainty associated with the parameter
$\kappa_\phi(P)$ or $\Lambda_P$.
 We then examine the effect of pseudoscalar poles
on the muon anomaly
based on this confidence level and
the values of $\Gamma(P \rightarrow \gamma\gamma)$
given in \cite{PDG}.

 The results of our calculation are 
summarized in Tables \ref{tab:vmd-pi}$-$\ref{tab:vmd2-etap}.
 They have all been obtained
by the integration routine VEGAS with
800 thousands sampling points per iteration
and iterated 15 times.
 From the explicit calculation,
the dependence on $\Gamma(P \rightarrow \gamma\gamma)$
and $\kappa_\phi(P)$ (or $\Lambda_P$)
is found to be rather simple,
as can be partly inferred from those tables.
 To obtain the most likely value and the error of $a_\mu (P)$
we follow the procedure described below
for the ${\rm VMD}_A$ model.
 First we take the average of $a_\mu (P)$ for
($\Gamma(P \rightarrow \gamma\gamma)_{\rm max}$,
 $\kappa_\phi(P)_{\rm center}$) 
and that for
($\Gamma(P \rightarrow \gamma\gamma)_{\rm min}$,
 $\kappa_\phi(P)_{\rm center}$) 
and deduce the errors associated with
$\Gamma(P \rightarrow \gamma\gamma)$ as the maximally 
allowed deviation from their mean value.
 Likewise the uncertainty associated with $\kappa_\phi (P)$
can be found from $a_\mu (P)$ for
($\Gamma(P \rightarrow \gamma\gamma)_{\rm center}$,
 $\kappa_\phi(P)_{\rm max}$) 
and that for
($\Gamma(P \rightarrow \gamma\gamma)_{\rm center}$,
 $\kappa_\phi(P)_{\rm min}$). 
 The central value of $a_\mu (P)$ is set equal 
to the mean value of the above two averages.
 The total uncertainty is obtained by taking the square 
root of the sum of the squares of the two errors, 
assuming that the errors associated with
$\Gamma(P \rightarrow \gamma\gamma)$ and $\kappa_\phi(P)$
are independently distributed.
 The same procedure also applies
to the ${\rm VMD}_B$ model
with ($\Gamma(P \rightarrow \gamma\gamma)$, $\Lambda(P)$) 
instead of
($\Gamma(P \rightarrow \gamma\gamma)$, $\kappa_\phi(P)$) .

 From the results listed
in Tables \ref{tab:vmd-pi}$-$\ref{tab:vmd-etap}
we find that
the contributions to the muon anomaly from $\pi^0$, $\eta^0$
and $\eta^\prime$ propagation in the ${\rm VMD}_A$ model
are given by
\begin{eqnarray}
 a_\mu(\pi^0,\,{\rm VMD}_A) &=&
 \displaystyle{
  -0.045\ 8\ (29) \times \left( \frac{\alpha}{\pi} \right)^3
 }, \nonumber \\
 a_\mu(\eta,\,{\rm VMD}_A) &=&
 \displaystyle{
  -0.010\ 6\ (9) \times \left( \frac{\alpha}{\pi} \right)^3
 }, \nonumber \\
 a_\mu(\eta^\prime,\,{\rm VMD}_A) &=&
 \displaystyle{
  -0.009\ 4\ (7) \times \left( \frac{\alpha}{\pi} \right)^3
 }. 
\end{eqnarray}
 For the ${\rm VMD}_B$ model we find similarly
\begin{eqnarray}
 a_\mu(\pi^0,\,{\rm VMD}_B) &=&
 \displaystyle{
  -0.045\ 8\ (29) \times \left( \frac{\alpha}{\pi} \right)^3
 }, \nonumber \\
 a_\mu(\eta,\,{\rm VMD}_B) &=&
 \displaystyle{
  -0.010\ 7\ (8) \times \left( \frac{\alpha}{\pi} \right)^3
 }, \nonumber \\
 a_\mu(\eta^\prime,\,{\rm VMD}_B) &=&
 \displaystyle{
  -0.009\ 5\ (7) \times \left( \frac{\alpha}{\pi} \right)^3
 }.
\end{eqnarray}
 The total contribution to muon $g-2$
from the pseudoscalar pole effect becomes
\begin{eqnarray}
 a_\mu(b,\,{\rm VMD}_A) &=&
 \displaystyle{
  -0.065\ 8\ (32) \times \left( \frac{\alpha}{\pi} \right)^3
 } \nonumber \\
 &=&
 \displaystyle{
  -82.5\ (4.1) \times 10^{-11}
 }, 
  \label{eq:vmd-1}
\end{eqnarray}
and
\begin{eqnarray}
 a_\mu(b,\,{\rm VMD}_B) &=&
 \displaystyle{
  -0.066\ 0\ (31) \times \left( \frac{\alpha}{\pi} \right)^3
 } \nonumber \\
 &=&
 \displaystyle{
  -82.7\ (3.9) \times 10^{-11}
 },
  \label{eq:vmd-2}
\end{eqnarray}
where the errors above were deduced assuming that 
the uncertainties associated with distinct mesons 
are normally distributed.
 These results 
are rather insensitive to the difference between
the two models, both of which are already strongly
constrained by the experiment.

 Let us comment here on the pion decay constant.
 Note that every postulated form factor used
in the previous analysis is proportional to $ 1/f_P $
implementing the PCAC.
 Thus the contribution to $ a_\mu $
from each pseudoscalar meson propagation
becomes proportional to $ 1/f_P^2 $.
 In the previous analysis \cite{K-had,HKS},
$f_\pi = 93$\,MeV was used.
 This is the decay constant of the charged pion.
 That of the neutral pion is quoted as
$ f_{\pi^0} \sim 84\ (3)$\,MeV in \cite{PDG}.
 Correspondingly
the values of $ a_\mu $ predicted by all the above
form factors receive the enhancement factor 1.22
if the latter value is used.
 But this implies
that too much isospin breaking (about 10 \%)
shows up in the pion decay constants
resulting from the mass difference of up and down quarks
and electromagnetic corrections.
 Those values found in Refs. \cite{CELLO,CLEO} for $\pi^0$
were extracted
from the use of pole-type form factor
(\ref{experimentalpiformfactor})
corresponding to the VMD picture
together with $\Lambda_{\pi^0}^2 = 8\pi^2f_{\pi^0}^2$.
 This relation must not be trusted
once we treat $\Gamma(P \rightarrow \gamma\gamma)$
and $\Lambda_P$ as two independent parameters.
 Its application might
lead to a fictitiously large isospin violation
in the pion decay constant.
 In our analysis throughout this paper,
the decay constants are not the direct input,
because its determination primarily depend
on the form factor assumed.
 Instead, we use the partial decay width
$\Gamma(P \rightarrow \gamma\gamma)$
given in \cite{PDG},
and determine the effective decay constant
according to an equation such as (\ref{eq:gamma-f}).
 This provides a phenomenological
and totally self-contained algorithm.
%

\section{Effect of the $P \gamma^* \gamma^*$ form factor
on muon $\lowercase{g-2}$}
\label{sec:formfactor}

 For a full evaluation of the muon anomaly $a_\mu (b)$,
we need information
on the form factor for all values of $p_1$ and $p_2$.
 In other words,
we need information
not only for large $p_1^2$ with $p_2^2 = 0$
or large $p_2^2$ with $p_1^2 = 0$
but also for the case where both $p_1^2$ and $p_2^2$
are large simultaneously.
 In the VMD model ($b_1$)
the off-shell $\pi^0 \gamma^* \gamma^*$ form factor
for large $p_1^2$ and $p_2^2$ takes the form
\begin{equation}
  F_{\pi^0} (p_1^2 , p_2^2 ,0 ) \propto {1 \over {p^4 }} 
   ~~~~{\rm for}~~~~ p_1^2,~ p_2^2~ \simeq~ p^2 .
\label{VMDasym}
\end{equation}
 On the other hand the QTL model ($b_2$)
or a nonperturbative estimate
shows much slower behavior ($ \sim 1/p^2 $) as is seen from
(\ref{case2}) and (\ref{novikov}).

 This suggests that 
the $\pi^0 \gamma^* \gamma^*$ form factor 
may have hard components for large $p_1^2$ and $ p_2^2$.
 Taking the asymptotic form (\ref{novikov}) into account,
we may, for instance, choose the function
\begin{eqnarray}
 F_{M_X}^{\rm LN}(p_1^2,\, p_2^2) &=&
 \displaystyle{
  \frac{1}{2}
   \left(
    \frac{M_X^2}{M_X^2 - p_1^2} +
    \frac{M_X^2}{M_X^2 - p_2^2}
   \right)
 },
 \label{eq:hardcomp}
\end{eqnarray}
as a candidate of the hard component.
 Another possibility is to use (\ref{case2}) itself.
 With this in mind we consider two models:
\begin{description}
 \item{(I)}\ \ 
  Linear combination of the ${\rm VMD}_A$ model
  and the function (\ref{eq:hardcomp}),
 \item{(II)}
  Linear combination of the ${\rm VMD}_A$ model
  and the QTL model\,($b_2$).
\end{description}
\noindent
 We will not discuss the 
${\rm VMD}_B$ model since the difference between the
${\rm VMD}_B$ model 
and ${\rm VMD}_A$ model 
is small.
 Let us introduce a parameter $\beta_P$ to describe
the deviation from the ${\rm VMD}_A$ model.
 The best value of $\beta_P$ depends
on the species $P$ of pseudoscalar meson.
 To be more specific,
we write the scalar part
of the invariant amplitude (\ref{eq:inv-amp}) as
\begin{eqnarray}
 M_P(p_1^2,\,p_2^2,\,q^2) &=&
 \displaystyle{
  \frac{1}{f_P} \frac{\alpha}{\pi}
  \left[
    F_\rho^{\beta_P}(p_1^2,\,p_2^2,\,q^2)
    + \kappa_\phi(P)
    F_\phi^{\beta_P}(p_1^2,\,p_2^2,\,q^2)
  \right]
 },
 \label{eq:extended}
\end{eqnarray}
where, for the extended model (I),
\begin{eqnarray}
 F_V^{\beta_P} (p_1^2,\, p_2^2,\, q^2)
 &=&
 \displaystyle{
  (1-\beta_P) F_V^{\rm VMD}(p_1^2,\, p_2^2)
  + \beta_P\,F_{M_X(P)}^{\rm LN}(p_1^2,\, p_2^2)
 },
 \label{eq:(I)}
\end{eqnarray}
and, for the extended model (II),
\begin{eqnarray}
 F_V^{\beta_P} (p_1^2,\, p_2^2,\, q^2)
 &=&
 \displaystyle{
  (1-\beta_P) F_V^{\rm VMD}(p_1^2,\, p_2^2)
  + \beta_P\,F_V^{\rm QTL}(p_1^2,\, p_2^2,\, q^2)
 }.
 \label{eq:(II)}
\end{eqnarray}
 Here $F_V^{\rm VMD}(p_1^2,\, p_2^2)$
is the form factor (\ref{eq:VMD-form})
in ${\rm VMD}_A$ model,
$F_{M_X}^{\rm LN}(p_1^2,\, p_2^2)$
is given by (\ref{eq:hardcomp}) in which
$M_X(P)$ is assumed, for simplicity,
to take the same value for $\rho$ and $\phi$, and
\begin{eqnarray}
 \displaystyle{
  F_V^{\rm QTL}(p_1^2,\, p_2^2,\, q^2)
 } &=&
 \displaystyle{
  \left\{
    \begin{array}{cr}
      I_{m_u}(p_1^2,\, p_2^2,\, q^2) &
       \quad \quad {\rm for}\ V = \rho, \\
      I_{m_s}(p_1^2,\, p_2^2,\, q^2) &
       \quad \quad {\rm for}\ V = \phi .
    \end{array}
  \right.
 } 
\end{eqnarray}
$I_{m_u}$ is defined in (\ref{functionF0}).
$I_{m_s}$ takes account of the fact that $\phi$ meson
is dominated by the $s\bar{s}$ state.
 Again the ``effective'' decay constant $f_P$ in
(\ref{eq:extended})
does not have the usual meaning
and is determined in the same way
as in Sec. \ref{subsec:effectofpgg}.
 In the extended model (II), for instance,
for each meson $P$ and a given value of
$\Gamma(P \rightarrow \gamma\gamma)$,
the direct fit to the experimental data determines
the parameters $\beta_P$ and $\kappa_\phi(P)$,
which give the least value of $\chi^2$ according to
(\ref{eq:f-m}).
 Once the values of
$\Gamma(P \rightarrow \gamma\gamma)$,
$\kappa_\phi(P)$ and $\beta_P$ are fixed,
a relation similar to (\ref{eq:gamma-f}) leads to $f_P$.

 Note that the extended model (I) of (\ref{eq:(I)})
has an incorrect asymptotic behavior
for finite $p_1^2 $ or $p_2^2 $.
 Thus it must be regarded as an empirical formula designed
to fit the experimental data 
and applicable only up to the largest momentum transfer
of the experimental data.
 We use it nevertheless because of simplicity and
because its effect on the muon $g-2$ is small
as far as $\beta_P$ is small.
 The extended model (II)
is also not very satisfactory since the QTL part
has a logarithmic growth
which distorts the form factor
even for moderately large momentum transfers.
 Thus both must be treated with some caution.

\subsection{Extended model (I)}
\label{subsec:case-1}

 The extended model (I)
involves one additional parameter $M_X$,
and the best fitting values
of $\kappa_\phi(P)$ and $\beta_P$ change as $M_X$ changes
even for the same value of
$\Gamma(P \rightarrow \gamma\gamma)$.
 To avoid time-consuming and excessive analysis,
we first perform two parameter fitting for fixed $M_X$ and
$\Gamma(P \rightarrow \gamma\gamma)$
to determine a set ($\beta_P$, $\kappa_\phi(P)$)
that gives the best $\chi^2$.
 Next we determine the range of $\kappa_\phi$ allowed
at $1 \sigma$ confidence level
with $\beta_P$ fixed
to the best value obtained above, and {\it vice versa}.
 The result of such a fitting
is summarized
in Tables \ref{tab:al-pi-par}$-$\ref{tab:al-etap-par}.
 They show that $\beta_P$ is small in general and thus
does not favor
strong presence of hard component.
 Note that a large $M_X$ implies near $q$-independence
of the term:
\begin{equation}
 \frac{M_X^2}{M_X^2 - q^2} \rightarrow 1 \quad
  {\rm for}\ M_X^2 \rightarrow \infty\ {\rm with\ fixed}\ 
  q^2.
\end{equation}
 This means
that the formula (\ref{eq:(I)})
tends to a constant $\beta_P$
and hence gives rise to a divergence
of the muon $g - 2$ proportional to $\ln M_X$.
 Larger $M_X$ also leads to larger minimum of $\chi^2$.
 Thus we restrict $M_X$ to the values below 6 GeV,
and examine the effect of this term on the muon $g-2$.

 The computation of muon $g-2$ has been performed
for several sets of ($\beta$, $\kappa_\phi$),
in which either $\beta$ or $\kappa_\phi$ is
its best fitted value for given $M_X$
and $\Gamma(P \rightarrow \gamma\gamma)$.
 The results
listed in Tables \ref{tab:al-pi}$-$\ref{tab:al-etap}
for the extended model (I)
have been obtained by integration with
800 thousands sampling points per iteration
which is iterated 15 times.
 The errors generated by the numerical integration itself
are not explicitly stated
since they are all under 0.05 \%,
far below the errors arising
from the uncertainty in
$\Gamma(P \rightarrow \gamma\gamma)$.

 Let us first focus our attention on the case $M_X = M_\rho$ 
in Tables \ref{tab:al-pi}$-$\ref{tab:al-etap},
which is the lowest value of the scale $M_X$
characterizing the modification of the high energy behavior
in the present model.
 To deal with these data we adopt the same procedure
as in the case of the VMD model.
 For instance the uncertainty associated with the variation of
$\Gamma (P \rightarrow \gamma \gamma )$
can be read off from $a_\mu (P)$ for
($\Gamma(P \rightarrow \gamma\gamma)_{\rm max}$,
 $\beta (P)_{\rm center}$, $\kappa_\phi(P)_{\rm center}$) 
and that for
($\Gamma(P \rightarrow \gamma\gamma)_{\rm min}$,
 $\beta (P)_{\rm center}$, $\kappa_\phi(P)_{\rm center}$).
 The uncertainty associated with the variation of 
$\beta (P)$ and $\kappa_\phi(P)$
will be found similarly.
 We take the average of three tentative central values to 
obtain the central value of $a_\mu (P)$ given below.
 The combined error is obtained by taking the square root of 
the sum of squares of the errors.
 The contributions to muon anomaly from $\pi^0$, $\eta^0$
and $\eta^\prime$ propagation in the extended model (I)
can then be written as
\begin{eqnarray}
 a_\mu(\pi^0,\,({\rm I})) &=&
 \displaystyle{
  -0.045\ 6\ (28) \times \left( \frac{\alpha}{\pi} \right)^3
 }, \nonumber \\
 a_\mu(\eta,\,({\rm I})) &=&
 \displaystyle{
  -0.010\ 3\ (9) \times \left( \frac{\alpha}{\pi} \right)^3
 }, \nonumber \\
 a_\mu(\eta^\prime,\,({\rm I})) &=&
 \displaystyle{
  -0.009\ 0\ (6) \times \left( \frac{\alpha}{\pi} \right)^3
 },
  \label{eq:ext-1}
\end{eqnarray}
where the uncertainties are estimated assuming
that the error distribution associated
with $\Gamma(P \rightarrow \gamma\gamma)$, $\beta_P$
and $\kappa_\phi(P)$ are normal and independent.
 The result obtained assuming $M_X$ = 2 GeV is nearly identical
with (\ref{eq:ext-1}).
 The result for larger values of $M_X$ deviates only slightly
from (\ref{eq:ext-1}),
except for the considerable $M_X$ dependence
seen in $a_\mu (\eta,\,({\rm I}))$.

\subsection{Extended model (II)}
\label{subsec:case-2}

 The trial form factor for the extended model (II) is
the ${\rm VMD}_A$ model
augmented slightly by a QTL term.
 The additional term
receives a strong constraint from the experimental data
as is shown in Tables \ref{tab:aq-pi-par}$-$\ref{tab:aq-etap-par}.
 This is anticipated
because the logarithmic factor in 
(\ref{case2}) of the QTL model
is quite visible even 
in the moderately large momentum region,
and hence receives a strong constraint from experiment.
 Tables \ref{tab:aq-pi}$-$\ref{tab:aq-etap} show
the prediction for the muon $g-2$
from the trial form factor (\ref{eq:(II)})
for each pseudoscalar meson
under the constraint of the parameters
in Tables \ref{tab:aq-pi-par}$-$\ref{tab:aq-etap-par},
in which all the results
have been obtained by the integration with
500 thousands sampling points per iteration
which is iterated 25 times.
 In the manner similar
to that of Sec. \ref{subsec:case-1},
the contributions to muon anomaly from $\pi^0$, $\eta^0$
and $\eta^\prime$ propagation in the extended model (II)
are estimated to be
\begin{eqnarray}
 a_\mu(\pi^0,\,({\rm II})) &=&
 \displaystyle{
  -0.045\ 6\ (28) \times \left( \frac{\alpha}{\pi} \right)^3
 }, \nonumber \\
 a_\mu(\eta,\,({\rm II})) &=&
 \displaystyle{
  -0.006\ 4\ (9) \times \left( \frac{\alpha}{\pi} \right)^3
 }, \nonumber \\
 a_\mu(\eta^\prime,\,({\rm II})) &=&
 \displaystyle{
  -0.008\ 7\ (7) \times \left( \frac{\alpha}{\pi} \right)^3
 }.
  \label{eq:ext-2}
\end{eqnarray}

 It is seen that
the QTL modification to the VMD results
of Sec. \ref{sec:comparison}
is larger compared with the case of extended model (I).
 In particular, the $\eta$ contribution is reduced in magnitude
compared to that obtained in the other scheme.
 Comparison of (\ref{eq:ext-2}) with (\ref{eq:ext-1})
shows that
the additional term in case (II) leads to
a better $\chi^2$.
 Moreover
the signs of $\kappa_\phi$ for $\eta$ and $\eta^\prime$
tend to become negative in the present case
compared to the ${\rm VMD}_A$ model,
as can be seen from Tables
\ref{tab:vmd-eta}, \ref{tab:vmd-etap},
\ref{tab:aq-eta-par} and \ref{tab:aq-etap-par}. 
 The opposite signs of $\rho$ and $\phi$
contributions result in the decrease of absolute magnitudes
of $\eta$ and $\eta^\prime$ contributions.
 This relative minus sign is required
to fit to the experimental data
once the form factor contains
a QTL component
which enhances the contribution of large momentum region.
 In the sense that 
this region is beyond control of the experimental data
and the logarithmic enhancement may be an
artifact of perturbation theory, however,
the result (\ref{eq:ext-2})
must be treated with some caution,
and should be accorded less weight
than the extended model (I). 

 Note that the flavor SU(3) relation implies
$\kappa_\phi(\eta) \simeq -0.28$
and $\kappa_\phi(\eta^\prime) \simeq 0.29$
for $\theta \simeq -10.1^{\,\circ}$ \cite{PDG}.
 While the signs of $\kappa_\phi(\eta)$ 
and $\kappa_\phi(\eta^\prime)$ in the extended model (I)
are consistent with the SU(3) relation,
those in the extended model (II)
are not.
 Of course this does not mean that
one model is more appropriate than the other.
 In the first place,
it is obscure
whether SU(3) breaking effect due to
strange quark mass is substantial or not.
 Secondly, we do not know
the relevance of the ${\rm U(1)}_A$ anomaly (QCD anomaly)
contribution
to the $\eta_0 \rightarrow \gamma\gamma$
($\eta_0$ is the SU(3) singlet component) vertex.
 These questions have been examined
for $P \rightarrow \gamma\gamma$
in  \cite{Shore},
and for another decay channel in \cite{QCD-anomaly},
in which the subtlety 
that the presence of QCD anomaly
raises is explored in identifying the decay constant
invariant under renormalization group,
inducing a nonperturbative ambiguity
associated with each decay channel.
 We shall not discuss its relevance quantitatively here.
 We believe 
that these effects can be incorporated
in a few constants parametrizing
the form factor we chose,
by fitting to the observed pseudoscalar production
cross section through the two photon process
\cite{CELLO,CLEO,CLEO-2}.

\section{Summary and Discussion}
\label{sec:discussion}

 In this paper
we reexamined the pseudoscalar meson pole contributions
to the muon $g-2$ taking account
of the measured $P \gamma \gamma^*$ form factor 
and of possible effect
of the postulated $P \gamma^* \gamma^*$ form factor.
 Insofar as we demand that these form factors
respect the measured strength of $P\gamma\gamma^*$
for large momentum ($\sim$  several GeV)
with one photon on the mass-shell,
the form factor cannot deviate from
the VMD model substantially.

 The results (\ref{eq:ext-1}) and (\ref{eq:ext-2})
of the extended models
have somewhat smaller overall uncertainties than the results 
(\ref{eq:vmd-1}) and (\ref{eq:vmd-2}) of the $\rm{VMD}_A$
and $\rm{VMD}_B$ models.
 This may be due to better flexibility of the extended models.

 Inspection of (\ref{eq:ext-1}) and (\ref{eq:ext-2}) shows that
the contribution of the $\pi^0 \gamma^* \gamma^*$ form factor
is nearly identical for both models
while the contribution
of the $\eta \gamma^* \gamma^*$ form factor
shows considerable dependence
on the assumed structure of hard component.
 This may not be surprising since the effect of hard component
will be stronger for $\eta$ than for $\pi^0$.
 Better agreement of the $\eta^\prime \gamma^* \gamma^*$ cases
may just be a matter of coincidence.
 The $\eta^\prime$ contribution in (\ref{eq:ext-1})
is somewhat smaller
than that of $\eta$,
in contrast to the results of the ENJL model
and the extended model (II),
both of which contain the QTL form factor 
at least partly.

 Note that either (\ref{eq:ext-1}) or (\ref{eq:ext-2})
cannot be taken as 
our best estimate based on the experimental data only since
they depend on theoretical assumption about the hard component.
 Instead they must be regarded as providing a measure
of uncertainty in $a_\mu$ due to the unknown effect
of the $P\gamma^* \gamma^*$ form factor.
 Our calculation shows that this uncertainty comes mainly from
the contribution of $\eta \gamma^* \gamma^*$ form factor.
 For the reasons discussed in Sec. \ref{subsec:case-2},
we believe that
$a_\mu (\eta,\,(\rm{II}))$ is less reliable than
$a_\mu (\eta,\,(\rm{I}))$.
 But their difference $0.0039\,(\alpha /\pi )^3$ may be
regarded as a measure of theoretical uncertainty.

 Based on these considerations, we adopt (\ref{eq:vmd-2})
as our best estimate and choose as its uncertainty
the statistical combination of the error in (\ref{eq:vmd-2}),
which comes from the experimental uncertainty 
in the $P\gamma \gamma^*$ measurement,
and the effect described above due to possible presence
of hard component in the $P\gamma^* \gamma^*$ form factor. 
 We thus arrive at the total pseudoscalar pole contribution 
\begin{equation}
 a_\mu(\pi + \eta + \eta^\prime) = 
  - 82.7\ (6.4) \times 10^{-11}.
 \label{eq:b-pseudo}
\end{equation}
 The uncertainty here has been reduced to one half of 
the previous result (\ref{HKSabc}).
 That the result (\ref{BPPabc}) from Ref. \cite{Bijnens}
is close to our result
(\ref{eq:b-pseudo})
is not surprising 
since the modified ENJL model 
of \cite{Bijnens} 
contains an adjustable free parameter.
 The crucial difference between
 the old results (\ref{HKSabc}), (\ref{BPPabc})
and the new result (\ref{eq:b-pseudo}) is 
that the latter is much less dependent
on the theoretical ambiguity, being determined 
by the experimental data to a large extent.

 The theoretical uncertainty in (\ref{eq:b-pseudo})
reflects the absence of
data on the $P\gamma^* \gamma^*$ form factor.
 It is useful to note, however, that
it is possible to gain some insight for
the effect of the $P \gamma^* \gamma^*$ form factor based on
the measurements of $\pi^0 \rightarrow e^+ e^-$ 
and $\eta \rightarrow \mu^+ \mu^-$ decays,
both of which involve the $P \gamma^* \gamma^*$ form factor.
 The measured branching ratios \cite{kessler}
\begin{eqnarray}
 B_{\rm meas} ( \eta \rightarrow \mu^+ \mu^- )  \equiv
 {{ \Gamma (\eta \rightarrow \mu^+ \mu^- )}  \over
 { \Gamma (\eta \rightarrow \gamma \gamma )}} =
 1.4\ (0.2) \times 10^{-5} ,
 \label{eq:etadecay}
\end{eqnarray}
and
\begin{eqnarray}
 B_{\rm meas} ( \pi^0 \rightarrow e^+ e^- )  \equiv
 {{ \Gamma (\pi^0 \rightarrow e^+ e^- )}  \over
 { \Gamma (\pi^0 \rightarrow \gamma \gamma )}} =
 7.3\ (1.9) \times 10^{-8} , 
  \label{eq:pi0decay}
\end{eqnarray}
where the latter is a weighted average of the data from
Ref. \cite{deshpande} are in good agreement
with the theoretical values based on the VMD model
\cite{ametller}
\footnote{We thank J. F. Donoghue 
for calling our attention to this reference.}
\begin{eqnarray}
 B_{\rm VMD} ( \eta \rightarrow \mu^+ \mu^- ) &\equiv&
 \displaystyle{
  {{ \Gamma (\eta \rightarrow \mu^+ \mu^- )} \over
   { \Gamma (\eta \rightarrow \gamma \gamma )}}
 } =
 (1.14^{+ 0.07}_{-0.03}) \times 10^{-5} , 
\nonumber  \\
 B_{\rm VMD} ( \pi^0 \rightarrow e^+ e^- ) &\equiv&
 \displaystyle{
  {{ \Gamma (\pi^0 \rightarrow e^+ e^- )} \over
   { \Gamma (\pi^0 \rightarrow \gamma \gamma )}}
 } =
  6.41\ (0.19) \times 10^{-8} .
   \label{eq:th.decay}
\end{eqnarray}
 The differences between these values are presumably due to
the effect of the $P\gamma^* \gamma^*$ form factor.
 It appears to be no greater than 15 \%.
 In principle, such an information can lead to 
a completely model-independent evaluation
of the pseudoscalar pole contribution
to the muon $g-2$.
 Further improvement in these measurements
will thus be of great interest.

 Unlike the pseudoscalar pole contribution
discussed in this paper,
the other contributions of hadronic light-by-light
scattering type from Fig. \ref{fig:diagram}
remain vulnerable to various theoretical ambiguities.
 The axial-vector pole contribution (\ref{BPPaxsc})
was obtained
entirely based on the ENJL model,
which is certainly not satisfactory in the sense that it
does not lead to the correct asymptotic behavior
for the $\pi^0 \gamma \gamma^*$ vertex \cite{Bijnens}.
 Based on the estimates 
(\ref{HKSaxial})
and (\ref{BPPaxsc}) 
the axial-vector pole contribution
seems to be relatively minor
compared with the effect (\ref{eq:b-pseudo})
of the pseudoscalar pole propagation.
 Since nothing better is available at present
and since it is very small, 
we choose (\ref{HKSaxial})
as our best estimate of the axial vector 
contribution assigning an uncertainty as large as
the value (\ref{HKSaxial}) itself.
 The axial-vector meson contribution
may be estimated more reliably if
the off-shell structure of $A\gamma\gamma$ vertex
is available from the experiment
as was the case for the $P \gamma \gamma $ vertex.

 Another possible pole effect, the scalar resonance
inferred within the ENJL model, has not been
observed convincingly.
 It may not be surprising if an "exact" QCD calculation leads
only to a continuum spectrum
instead of producing a broad resonance
in the scalar channel of the corresponding energy scale
( $0.6 \lesssim \sqrt{s} \lesssim 1.5$\ GeV ).

 In the absence of any resonance in the low energy region
(below several hundred MeV)
in the $\pi -\pi$ and $q \bar{q}$ channels,
the pion-loop contribution $a_\mu (a)$ and
the quark-loop contribution $a_\mu (c)$ may be regarded as
representing the effect of the continuum spectrum.
 Previous analysis \cite{HKS,Bijnens} shows that
$a_\mu (a)$ and $a_\mu (c)$ have sizable contributions
from the region of higher momentum transfer where
neither chiral perturbation theory
nor the $1/N_c$ expansion may provide reliable guidance.
 Even mutual independence of the three types of
diagrams of Fig. \ref{fig:diagram} might not hold valid there.
 For lack of better argument, however,
we choose the sum of $a_\mu (a)$ and $a_\mu (c)$
as a crude estimate of the entire continuum
contribution.
 The scalar pole contribution may be regarded as
included in this estimate.

 Finally,
collecting $a_\mu(a)$ and $a_\mu(c)$ from (\ref{HKSabc}),
the pseudoscalar meson pole contribution 
from (\ref{eq:b-pseudo}),
and including (\ref{HKSaxial}),
we present 
\begin{equation}
 a_\mu(\mbox{had l-l}) = - 79.2\ (15.4) \times 10^{-11} 
\label{hadlbyl}
\end{equation}
as our best estimate of the total
hadronic light-by-light scattering contribution
to the muon anomaly.
 Here the uncertainty is obtained from
the errors of various components
assuming normal distribution.
 If the uncertainties are combined additively,
one would obtain $\sim \pm 30 \times 10^{-11}$.

 Further progress in theory would have to wait
for the lattice QCD calculation of the hadronic
four-point function.
 In view of the recent progress in the lattice QCD
and rapidly increasing computing power,
such a calculation may no longer be a far-fetched dream.
 As of now, the theoretical value of the muon $g-2$,
which consists of (\ref{amuQED}), (\ref{had.v.p. contr}),
(\ref{higherhad.v.p. contr}), (\ref{weakeffect}), and
our new analysis
of the hadronic light-by-light scattering effect
(\ref{hadlbyl}),
is given by
\begin{equation}
 a_\mu ({\rm th}) = 116\ 591\ 687\ (96) \times 10^{-11}.
 \label{eq:newprediction}
\end{equation}
 The largest source of theoretical uncertainty remains
to be the hadronic vacuuum polarization effect
(\ref{had.v.p. contr}).
 Better measurements of this effect are urgently needed.

\acknowledgements

 T. K. thanks the hospitality of KEK,
High Energy Accelerator Research Organization,
where part of this work was carried out.
 We thank V. Savinov for informing us of
the up-dated result from CLEO and for useful discussions.
 Thanks are due to B. N. Taylor, P. J. Mohr, and B. L. Roberts
for helpful comments.
 T. K.'s work is supported
in part by the U. S. National Science Foundation. 
 M. H. is a JSPS Fellow supported in part
by the Grand-in-Aid for Scientific Research
from the Ministry of Education, Science and Culture
of Japan.

%

%
%
\begin{table}
 \caption{ $\pi^0$ contribution to muon $g-2$
           in the ${\rm VMD}_A$ model.}
 \label{tab:vmd-pi}
 \vspace{0.3cm}
 \begin{tabular}{crc}
   $\Gamma(\pi^0 \rightarrow \gamma\gamma)$\ [eV]
   & $\kappa_\phi(\pi^0)$
   & $a_\mu(\pi^0) / \left(\alpha/\pi\right)^3$ \\
   \hline
   7.22 & $-$0.01 & $-$0.041\ 87\ (2) \\
   \hline
   7.22 & 0.12 & $-$0.043\ 56\ (2) \\
   \hline
   7.22 & 0.27  & $-$0.045\ 28\ (2) \\
   \hline
   7.78 & $-$0.07  & $-$0.044\ 03\ (2) \\
   \hline
   7.78 & 0.04  & $-$0.045\ 80\ (2) \\
   \hline
   7.78 & 0.16  & $-$0.047\ 54\ (2) \\ 
   \hline
   8.34 & $-$0.12  & $-$0.046\ 17\ (2) \\
   \hline
   8.34 & $-$0.03 & $-$0.047\ 95\ (2) \\
   \hline
   8.34 & 0.08  & $-$0.049\ 76\ (2) \\
 \end{tabular}
\end{table}
\begin{table}
 \caption{ $\eta$ contribution to muon $g-2$
           in the ${\rm VMD}_A$ model.}
 \label{tab:vmd-eta}
 \vspace{0.3cm}
 \begin{tabular}{crc}
   $\Gamma(\eta \rightarrow \gamma\gamma)$\ [keV]
   & $\kappa_\phi(\eta)$
   & $a_\mu(\eta) / \left(\alpha/\pi\right)^3$ \\
   \hline
   0.42 & 0.04    & $-$0.009\ 646\ (4) \\
   \hline
   0.42 & 0.19    & $-$0.010\ 235\ (4) \\
   \hline
   0.42 & 0.38    & $-$0.010\ 843\ (4) \\
   \hline
   0.46 & $-$0.04 & $-$0.009\ 910\ (4) \\
   \hline
   0.46 & 0.07    & $-$0.010\ 726\ (4) \\
   \hline
   0.46 & 0.22    & $-$0.011\ 349\ (4) \\ 
   \hline
   0.50 & $-$0.10 & $-$0.010\ 594\ (4) \\
   \hline
   0.50 & $-$0.01 & $-$0.011\ 211\ (4) \\
   \hline
   0.50 & 0.13    & $-$0.011\ 921\ (5) \\
 \end{tabular}
\end{table}
\begin{table}
 \caption{ $\eta^\prime$ contribution to muon $g-2$
           in the ${\rm VMD}_A$ model.}
 \label{tab:vmd-etap}
 \vspace{0.3cm}
 \begin{tabular}{crc}
   $\Gamma(\eta^\prime \rightarrow \gamma\gamma)$\ [keV] &
   $\kappa_\phi(\eta^\prime)$ &
   $a_\mu(\eta^\prime) / \left(\alpha/\pi\right)^3$ \\
   \hline
   4.07 & 0.25 & $-$0.008\ 584\ (4) \\
   \hline
   4.07 & 0.48 & $-$0.009\ 190\ (4) \\
   \hline
   4.07 & 0.82 & $-$0.009\ 818\ (5) \\
   \hline
   4.26 & 0.19 & $-$0.008\ 766\ (4) \\
   \hline
   4.26 & 0.39 & $-$0.009\ 380\ (4) \\
   \hline
   4.26 & 0.67 & $-$0.010\ 013\ (5) \\ 
   \hline
   4.45 & 0.13 & $-$0.008\ 947\ (4) \\
   \hline
   4.45 & 0.31 & $-$0.009\ 562\ (4) \\
   \hline
   4.45 & 0.55 & $-$0.010\ 205\ (5) \\
 \end{tabular}
\end{table}
\begin{table}
 \caption{ $\pi^0$ contribution to muon $g-2$ in
         the ${\rm VMD}_B$ model.}
 \label{tab:vmd2-pi}
 \vspace{0.3cm}
 \begin{tabular}{ccc}
   $\Gamma(\pi^0 \rightarrow \gamma\gamma)$\ [eV]
   & $\Lambda_{\pi^0}$ \ [GeV]
   & $a_\mu(\pi^0) / \left(\alpha/\pi\right)^3$ \\
   \hline
   7.22 & 0.77 & $-$0.041\ 86\ (2) \\
   \hline
   7.22 & 0.80 & $-$0.043\ 78\ (2) \\
   \hline
   7.22 & 0.82 & $-$0.045\ 67\ (2) \\
   \hline
   7.78 & 0.75 & $-$0.043\ 83\ (2) \\
   \hline
   7.78 & 0.78 & $-$0.045\ 85\ (2) \\
   \hline
   7.78 & 0.80 & $-$0.047\ 83\ (2) \\ 
   \hline
   8.34 & 0.73 & $-$0.045\ 75\ (2) \\
   \hline
   8.34 & 0.76 & $-$0.047\ 86\ (2) \\
   \hline
   8.34 & 0.79 & $-$0.049\ 93\ (2) \\
 \end{tabular}
\end{table}
\begin{table}
 \caption{ $\eta$ contribution to muon $g-2$ in
         the  ${\rm VMD}_B$ model.}
 \label{tab:vmd2-eta}
 \vspace{0.3cm}
 \begin{tabular}{ccc}
   $\Gamma(\eta \rightarrow \gamma\gamma)$\ [keV]
   & $\Lambda_\eta$\ [GeV]
   & $a_\mu(\eta) /  \left(\alpha/\pi\right)^3$ \\
   \hline
   0.42 & 0.78 & $-$0.009\ 645\ (4) \\
   \hline
   0.42 & 0.81 & $-$0.010\ 291\ (4) \\
   \hline
   0.42 & 0.84 & $-$0.010\ 940\ (4) \\
   \hline
   0.46 & 0.76 & $-$0.010\ 066\ (4) \\
   \hline
   0.46 & 0.79 & $-$0.010\ 742\ (4) \\
   \hline
   0.46 & 0.81 & $-$0.011\ 419\ (5) \\ 
   \hline
   0.50 & 0.74 & $-$0.010\ 467\ (4) \\
   \hline
   0.50 & 0.77 & $-$0.011\ 171\ (4) \\
   \hline
   0.50 & 0.79 & $-$0.011\ 876\ (5) \\
 \end{tabular}
\end{table}
\begin{table}
 \caption{ $\eta^\prime$ contribution to muon $g-2$ in
          the ${\rm VMD}_B$ model.}
 \label{tab:vmd2-etap}
 \vspace{0.3cm}
 \begin{tabular}{ccc}
   $\Gamma(\eta^\prime \rightarrow \gamma\gamma)$\ [keV] &
   $\Lambda_{\eta^\prime}$\ [GeV] &
   $a_\mu(\eta^\prime) / \left(\alpha/\pi\right)^3$ \\
   \hline
   4.07 & 0.82 & $-$0.008\ 655\ (4) \\
   \hline
   4.07 & 0.85 & $-$0.009\ 295\ (4) \\
   \hline
   4.07 & 0.88 & $-$0.009\ 942\ (5) \\
   \hline
   4.26 & 0.81 & $-$0.008\ 823\ (4) \\
   \hline
   4.26 & 0.84 & $-$0.009\ 475\ (4) \\
   \hline
   4.26 & 0.87 & $-$0.010\ 135\ (5) \\ 
   \hline
   4.45 & 0.80 & $-$0.008\ 986\ (4) \\
   \hline
   4.45 & 0.83 & $-$0.009\ 651\ (5) \\
   \hline
   4.45 & 0.86 & $-$0.010\ 323\ (5) \\
 \end{tabular}
\end{table}
%
%
%

\begin{table}
 \caption{ Values of $\kappa_\phi$ and $\beta$ 
           in the extended model (I)
           for $\pi^0$
           for the various values of $M_X$ and
           $\Gamma(\pi^0 \rightarrow \gamma\gamma)$.
           The error accompanying each parameter is
           obtained with the other parameter fixed
           to the best fitted value
           and at $1 \sigma$ confidence level.
         }
 \label{tab:al-pi-par}
 \vspace{0.3cm}
 \begin{tabular}{cccc}
   $M_X$\ [GeV] &
   $\Gamma(\pi^0 \rightarrow \gamma\gamma)$\ [eV] &
   $\kappa_\phi(\pi^0)$ &
   $\beta(\pi^0)$ \\
   \hline
   $M_\rho$ & 7.22 &
   \ \ \,$0.12_{-0.12}^{+0.16}$ &
   $-0.0003_{-0.0218}^{+0.0215}$ \\
   \hline
   $M_\rho$ & 7.78 &
   \ \ \,$0.02_{-0.10}^{+0.12}$ &
   \ \ \,$0.0035_{-0.0204}^{+0.0204}$ \\
   \hline
   $M_\rho$ & 8.34 &
   $-0.06_{-1.35}^{+1.01}$ &
   \ \ \,$0.0065_{-0.0194}^{+0.0194}$ \\
   \hline
   2.0 & 7.22 &
   \ \ \,$0.12_{-0.12}^{+0.15}$ &
   $-0.0005_{-0.0150}^{+0.0150}$ \\
   \hline
   2.0 & 7.78 &
   \ \ \,$0.02_{-0.10}^{+0.12}$ &
   \ \ \,$0.0024_{-0.0143}^{+0.0143}$ \\
   \hline
   2.0 & 8.34 &
   $-0.06_{-0.08}^{+0.10}$ &
   \ \ \,$0.0055_{-0.0137}^{+0.0136}$ \\ 
   \hline
   4.0 & 7.22 &
   \ \ \,$0.12_{-0.12}^{+0.16}$ &
   $-0.0004_{-0.0121}^{+0.0121}$ \\
   \hline
   4.0 & 7.78 &
   \ \ \,$0.02_{-0.10}^{+0.10}$ &
   \ \ \,$0.0020_{-0.0116}^{+0.0115}$ \\
   \hline
   4.0 & 8.34 &
   $-0.06_{-0.08}^{+0.12}$ &
   \ \ \,$0.0040_{-0.0111}^{+0.0111}$ \\ 
   \hline
   6.0 & 7.22 &
   \ \ \,$0.12_{-0.12}^{+0.16}$ &
   $-0.0003_{-0.0114}^{+0.0114}$ \\
   \hline
   6.0 & 7.78 &
   \ \ \,$0.02_{-0.10}^{+0.12}$ &
   \ \ \,$0.0018_{-0.0108}^{+0.0108}$ \\
   \hline
   6.0 & 8.34 &
   $-0.06_{-0.08}^{+0.10}$ &
   \ \ \,$0.0036_{-0.0104}^{+0.0104}$ \\ 
 \end{tabular}
\end{table}
\begin{table}
 \caption{ Values of $\kappa_\phi$ and $\beta$ 
           in the extended model (I) 
           for $\eta$.
           The meaning of numbers are the same as
           in Table {\protect \ref{tab:al-pi-par}}.
          }
 \label{tab:al-eta-par}
 \vspace{0.3cm}
 \begin{tabular}{cccc}
   $M_X$ [GeV] &
   $\Gamma(\eta \rightarrow \gamma\gamma)$\ [keV] &
   $\kappa_\phi(\eta)$ &
   $\beta(\eta)$ \\
   \hline
   $M_\rho$ & 0.42 &
   $-0.12_{-0.09}^{+0.11}$ &
   \ \ $0.059_{-0.022}^{+0.022}$ \\
   \hline
   $M_\rho$ & 0.46 &
   $-0.19_{-0.07}^{+0.09}$ &
   \ \ $0.060_{-0.020}^{+0.020}$ \\
   \hline
   $M_\rho$ & 0.50 &
   $-0.25_{-0.06}^{+0.07}$ &
   \ \ $0.061_{-0.019}^{+0.019}$ \\
   \hline
   2.0 & 0.42 &
   $-0.19_{-0.07}^{+0.09}$ &
   \ \ $0.053_{-0.015}^{+0.015}$ \\
   \hline
   2.0 & 0.46 &
   $-0.25_{-0.06}^{+0.07}$ &
   \ \ $0.054_{-0.014}^{+0.014}$ \\
   \hline
   2.0 & 0.50 &
   $-0.29_{-0.05}^{-0.06}$ &
   \ \ $0.054_{-0.014}^{+0.014}$ \\ 
   \hline
   4.0 & 0.42 &
   $-0.16_{-0.08}^{+0.10}$ &
   \ \ $0.040_{-0.013}^{+0.013}$ \\
   \hline
   4.0 & 0.46 &
   $-0.23_{-0.06}^{+0.08}$ &
   \ \ $0.040_{-0.012}^{+0.012}$ \\
   \hline
   4.0 & 0.50 &
   $-0.27_{-0.05}^{+0.06}$ &
   \ \ $0.041_{-0.011}^{+0.011}$ \\ 
   \hline
   6.0 & 0.42 &
   $-0.15_{-0.08}^{+0.10}$ &
   \ \ $0.035_{-0.012}^{+0.012}$ \\
   \hline
   6.0 & 0.46 &
   $-0.21_{-0.07}^{+0.08}$ &
   \ \ $0.035_{-0.011}^{+0.011}$ \\
   \hline
   6.0 & 0.50 &
   $-0.26_{-0.06}^{+0.07}$ &
   \ \ $0.036_{-0.011}^{+0.011}$ \\ 
 \end{tabular}
\end{table}
\begin{table}
 \caption{ Values of $\kappa_\phi$
          and $\beta$ 
          in the extended model (I) 
          for $\eta^\prime$.
           The meaning of numbers are the same as
          in Table {\protect \ref{tab:al-pi-par}}.
          }
 \label{tab:al-etap-par}
 \vspace{0.3cm}
 \begin{tabular}{cccc}
   $M_X$ [GeV] &
   $\Gamma(\eta^\prime \rightarrow \gamma\gamma)$\ [keV] &
   $\kappa_\phi(\eta^\prime)$ &
   $\beta(\eta^\prime)$ \\
   \hline
   $M_\rho$ & 4.07 &
   \ \ \,$0.30_{-0.18}^{+0.25}$ &
   \ \ \,$0.015_{-0.017}^{+0.017}$ \\
   \hline
   $M_\rho$ & 4.26 &
   \ \ \,$0.22_{-0.16}^{+0.22}$ &
   \ \ \,$0.016_{-0.016}^{+0.016}$ \\
   \hline
   $M_\rho$ & 4.45 &
   \ \ \,$0.15_{-0.14}^{+0.19}$ &
   \ \ \,$0.017_{-0.016}^{+0.016}$ \\
   \hline
   2.0 & 4.07 &
   \ \ \,$0.25_{-0.17}^{+0.23}$ &
   \ \ \,$0.015_{-0.013}^{+0.013}$ \\
   \hline
   2.0 & 4.26 &
   \ \ \,$0.17_{-0.15}^{+0.20}$ &
   \ \ \,$0.016_{-0.013}^{+0.013}$ \\
   \hline
   2.0 & 4.45 &
   \ \ \,$0.11_{-0.13}^{+0.17}$ &
   \ \ \,$0.016_{-0.012}^{+0.012}$ \\ 
   \hline
   4.0 & 4.07 &
   \ \ \,$0.25_{-0.17}^{+0.23}$ &
   \ \ \,$0.012_{-0.010}^{+0.010}$ \\
   \hline
   4.0 & 4.26 &
   \ \ \,$0.17_{-0.15}^{+0.20}$ &
   \ \ \,$0.012_{-0.010}^{+0.010}$ \\
   \hline
   4.0 & 4.45 &
   \ \ \,$0.11_{-0.14}^{+0.17}$ &
   \ \ \,$0.013_{-0.010}^{+0.010}$ \\ 
   \hline
   6.0 & 4.07 &
   \ \ \,$0.27_{-0.17}^{+0.24}$ &
   \ \ \,$0.010_{-0.009}^{+0.009}$ \\
   \hline
   6.0 & 4.26 &
   \ \ \,$0.19_{-0.15}^{+0.20}$ &
   \ \ \,$0.010_{-0.009}^{+0.009}$ \\
   \hline
   6.0 & 4.45 &
   \ \ \,$0.13_{-0.13}^{+0.18}$ &
   \ \ \,$0.011_{-0.009}^{+0.009}$ \\ 
 \end{tabular}
\end{table}
%
%
\begin{table}
 \caption{ $\pi^0$ pole contribution to muon $g-2$
           in the extended model (I).
           The four numerals
           corresponding to
           one set of ($\beta$, $\kappa_\phi$) are
           the values of $a_\mu /(10^{-2} (\alpha/\pi)^3)$
           for the central value of
           $\Gamma(\pi^0 \rightarrow \gamma\gamma)$, and,
           from the above, for $M_X$ = $M_\rho$,
           2.0, 4.0 and 6.0 GeV respectively. 
           The errors accompanying them are inferred
           from the uncertainty in
           $\Gamma(\pi^0 \rightarrow \gamma\gamma)$.
         }
 \label{tab:al-pi}
 \vspace{0.3cm}
 \begin{tabular}{lccc}
    & $\beta(\pi^0)_{\rm min}$ &
   $\beta(\pi^0)_{\rm center}$ &
   $\beta(\pi^0)_{\rm max}$ \\
   \hline
  $\kappa_\phi(\pi^0)_{\rm min}$ &  &
    $\begin{array}{r}
      -4.391_{-0.207}^{+0.203} \\
      -4.402_{-0.220}^{+0.214} \\
      -4.407_{-0.226}^{+0.220} \\
      -4.411_{-0.225}^{+0.225}
     \end{array}$ &
  \\
  \hline
  $\kappa_\phi(\pi^0)_{\rm center}$ &
   $\begin{array}{r}
      -4.498_{-0.185}^{+0.177} \\
      -4.442_{-0.192}^{+0.186} \\
      -4.417_{-0.199}^{+0.191} \\
      -4.400_{-0.202}^{+0.193}
     \end{array}$ &
   $\begin{array}{r}
      -4.559_{-0.208}^{+0.206} \\
      -4.570_{-0.219}^{+0.212} \\
      -4.576_{-0.224}^{+0.216} \\
      -4.579_{-0.229}^{+0.223}
    \end{array}$ &
   $\begin{array}{r}
      -4.621_{-0.231}^{+0.227} \\
      -4.701_{-0.245}^{+0.240} \\
      -4.736_{-0.252}^{+0.246} \\
      -4.761_{-0.257}^{+0.253}
    \end{array}$
   \\
   \hline
   $\kappa_\phi(\pi^0)_{\rm max}$ &  &
   $\begin{array}{r}
     -4.731_{-0.208}^{+0.201} \\
     -4.742_{-0.216}^{+0.211} \\
     -4.748_{-0.223}^{+0.218} \\
     -4.751_{-0.228}^{+0.221}
    \end{array}$ & \\
 \end{tabular}
\end{table}
\begin{table}
 \caption{ $\eta$ contribution to muon $g-2$
           in the extended model (I).
           The numbers
           corresponding to
           one set of ($\beta$, $\kappa_\phi$)
           have the meaning similar to those of
           Table {\protect \ref{tab:al-pi}}. 
         }
 \label{tab:al-eta}
 \vspace{0.3cm}
 \begin{tabular}{lccc}
    & $\beta(\eta)_{\rm min}$ & $\beta(\eta)_{\rm center}$ &
   $\beta(\eta)_{\rm max}$ \\
   \hline
  $\kappa_\phi(\eta)_{\rm min}$ &  &
    $\begin{array}{r}
       -0.995_{-0.062}^{+0.063} \\
       -1.119_{-0.083}^{+0.082} \\
       -1.154_{-0.086}^{+0.085} \\
       -1.174_{-0.087}^{+0.087}
     \end{array}$ &
   \\
   \hline
  $\kappa_\phi(\eta)_{\rm center}$ &
   $\begin{array}{r}
      -0.990_{-0.056}^{+0.055} \\
      -1.067_{-0.073}^{+0.072} \\
      -1.090_{-0.075}^{+0.074} \\
      -1.104_{-0.076}^{+0.075}
    \end{array}$ &
   $\begin{array}{r}
      -1.031_{-0.062}^{+0.063} \\
      -1.146_{-0.083}^{+0.083} \\
      -1.182_{-0.085}^{+0.086} \\
      -1.202_{-0.088}^{+0.087}
    \end{array}$ &
   $\begin{array}{r}
      -1.074_{-0.070}^{+0.071} \\
      -1.229_{-0.094}^{+0.094} \\
      -1.278_{-0.098}^{+0.098} \\
      -1.308_{-0.100}^{+0.101}
    \end{array}$ \\
   \hline
  $\kappa_\phi(\eta)_{\rm max}$ &  &
   $\begin{array}{r}
      -1.069_{-0.064}^{+0.064} \\
      -1.173_{-0.084}^{+0.083} \\
      -1.210_{-0.087}^{+0.086} \\
      -1.230_{-0.088}^{+0.088} \\
    \end{array}$ & \\
 \end{tabular}
\end{table}
\begin{table}
 \caption{ $\eta^\prime$ contribution to muon $g-2$
           in the extended model (I).
           The numbers corresponding to
           one set of ($\beta$, $\kappa_\phi$)
           have the meaning similar to those of
           Table {\protect \ref{tab:al-pi}}. 
         }
 \label{tab:al-etap}
 \vspace{0.3cm}
 \begin{tabular}{lccc}
    & $\beta(\eta^\prime)_{\rm min}$ &
   $\beta(\eta^\prime)_{\rm center}$ &
   $\beta(\eta^\prime)_{\rm max}$ \\
   \hline
  $\kappa_\phi(\eta^\prime)_{\rm min}$ &  &
    $\begin{array}{r}
       -0.842_{-0.018}^{+0.018} \\
       -0.859_{-0.021}^{+0.021} \\
       -0.872_{-0.021}^{+0.021} \\
       -0.879_{-0.022}^{+0.022} \\
    \end{array}$ & \\
   \hline
  $\kappa_\phi(\eta^\prime)_{\rm center}$ &
   $\begin{array}{r}
       -0.887_{-0.016}^{+0.016} \\
       -0.878_{-0.017}^{+0.017} \\
       -0.881_{-0.018}^{+0.018} \\
       -0.884_{-0.018}^{+0.018} \\
    \end{array}$ &
   $\begin{array}{r}
       -0.897_{-0.018}^{+0.018} \\
       -0.910_{-0.021}^{+0.021} \\
       -0.923_{-0.022}^{+0.022} \\
       -0.930_{-0.022}^{+0.022} \\
    \end{array}$ &
   $\begin{array}{r}
       -0.906_{-0.021}^{+0.021} \\
       -0.942_{-0.025}^{+0.025} \\
       -0.965_{-0.026}^{+0.026} \\
       -0.977_{-0.027}^{+0.027}
    \end{array}$ \\
   \hline
  $\kappa_\phi(\eta^\prime)_{\rm max}$ &  &
   $\begin{array}{r}
       -0.953_{-0.019}^{+0.019} \\
       -0.962_{-0.021}^{+0.021} \\
       -0.975_{-0.022}^{+0.022} \\
       -0.982_{-0.023}^{+0.023}
    \end{array}$ & \\
 \end{tabular}
\end{table}
\vfill
\newpage
%

\begin{table}
 \caption{ $\beta$ and $\kappa_\phi$ 
           in the extended model (II)
           for $\pi^0$
           for the various values of
           $\Gamma(\pi^0 \rightarrow \gamma\gamma)$.
           The error accompanying each parameter is
           obtained with the other parameter fixed
           to the best fitted value
           and at $1 \sigma$ confidence level.
         }
 \label{tab:aq-pi-par}
 \vspace{0.3cm}
 \begin{tabular}{ccc}
   $\Gamma(\pi^0 \rightarrow \gamma\gamma)$\ [eV] &
   $\kappa_\phi(\pi^0)$ &
   $\beta(\pi^0)$ \\
   \hline
   7.22 &
   \ \ \,$0.13_{-0.12}^{+0.16}$ &
   $-0.004_{-0.043}^{+0.042}$ \\
   \hline
   7.78 &
   \ \ \,$0.07_{-0.10}^{+0.12}$ &
   \ \ \,$0.011_{-0.043}^{+0.043}$ \\
   \hline
   8.34 &
   $-0.09_{-0.08}^{+0.09}$ &
   \ \ \,$0.028_{-0.044}^{+0.047}$ \\
 \end{tabular}
\end{table}
\begin{table}
 \caption{ $\beta$ and $\kappa_\phi$ 
           in the extended model (II) for $\eta$.
           The meaning of numbers is the same as
           in Table {\protect \ref{tab:aq-pi-par}}.
          }
 \label{tab:aq-eta-par}
 \vspace{0.3cm}
 \begin{tabular}{ccc}
   $\Gamma(\eta \rightarrow \gamma\gamma)$\ [keV] &
   $\kappa_\phi(\eta)$ & $\beta(\eta)$ \\
   \hline
   0.42 &
   $-0.28_{-0.05}^{+0.06}$ &
   \ \ $0.30_{-0.07}^{+0.08}$ \\
   \hline
   0.46 &
   $-0.30_{-0.05}^{+0.05}$ &
   \ \ $0.27_{-0.07}^{+0.08}$ \\
   \hline
   0.50 &
   $-0.32_{-0.04}^{+0.05}$ &
   \ \ $0.24_{-0.06}^{+0.07}$ \\
 \end{tabular}
\end{table}
\begin{table}
 \caption{ $\beta$ and $\kappa_\phi$ 
           in the extended model (II)
           for $\eta^\prime$.
           The meaning of numbers is the same as
           in Table {\protect \ref{tab:aq-pi-par}}.
          }
 \label{tab:aq-etap-par}
 \vspace{0.3cm}
 \begin{tabular}{ccc}
   $\Gamma(\eta^\prime \rightarrow \gamma\gamma)$\ [keV] &
   $\kappa_\phi(\eta^\prime)$ &
   $\beta(\eta^\prime)$ \\
   \hline
   4.07 &
   $-0.31_{-0.05}^{+0.05}$ &
   \ \ \,$0.20_{-0.03}^{+0.03}$ \\
   \hline
   4.26 &
   $-0.30_{-0.05}^{+0.05}$ &
   \ \ \,$0.19_{-0.03}^{+0.03}$ \\
   \hline
   4.45 &
   $-0.28_{-0.05}^{+0.05}$ &
   \ \ \,$0.17_{-0.03}^{+0.03}$ \\
 \end{tabular}
\end{table}
%
%
\begin{table}
 \caption{ $\pi^0$ pole contribution to
           $a_\mu/(10^{-2}(\alpha/\pi)^3)$
           in the extended model (II).
           The errors accompanied with them are inferred
           from the uncertainty in
           $\Gamma(\pi^0 \rightarrow \gamma\gamma)$.
         }
 \label{tab:aq-pi}
 \vspace{0.3cm}
 \begin{tabular}{lccc}
    & $\beta(\pi^0)_{\rm min}$ &
   $\beta(\pi^0)_{\rm center}$ &
   $\beta(\pi^0)_{\rm max}$ \\
   \hline
  $\kappa_\phi(\pi^0)_{\rm min}$ &  &
    $-4.386_{-0.183}^{+0.193}$ &
  \\
  \hline
  $\kappa_\phi(\pi^0)_{\rm center}$ &
   $-4.457_{-0.189}^{+0.186}$ &
   $-4.559_{-0.188}^{+0.196}$ &
   $-4.662_{-0.187}^{+0.196}$
   \\
   \hline
   $\kappa_\phi(\pi^0)_{\rm max}$ &  &
   $-4.736_{-0.191}^{+0.201}$& \\
 \end{tabular}
\end{table}
\begin{table}
 \caption{ $\eta$ pole contribution to
           $a_\mu/(10^{-2}(\alpha/\pi)^3)$
           in the extended model (II).
         }
 \label{tab:aq-eta}
 \vspace{0.3cm}
 \begin{tabular}{lccc}
    & $\beta(\eta)_{\rm min}$ & $\beta(\eta)_{\rm center}$ &
   $\beta(\eta)_{\rm max}$ \\
   \hline
  $\kappa_\phi(\eta)_{\rm min}$ &  &
    $-0.582_{-0.048}^{+0.047}$&
   \\
   \hline
  $\kappa_\phi(\eta)_{\rm center}$ &
   $-0.677_{-0.052}^{+0.053}$ &
   $-0.638_{-0.050}^{+0.049}$ &
   $-0.600_{-0.047}^{+0.046}$ \\
   \hline
  $\kappa_\phi(\eta)_{\rm max}$ &  &
   $-0.697_{-0.051}^{+0.051}$ & \\
 \end{tabular}
\end{table}
\begin{table}
 \caption{ $\eta^\prime$ pole contribution to
           $a_\mu/(10^{-2}(\alpha/\pi)^3)$
           in the extended model (II).
         }
 \label{tab:aq-etap}
 \vspace{0.3cm}
 \begin{tabular}{lccc}
    & $\beta(\eta^\prime)_{\rm min}$ &
   $\beta(\eta^\prime)_{\rm center}$ &
   $\beta(\eta^\prime)_{\rm max}$ \\
   \hline
  $\kappa_\phi(\eta^\prime)_{\rm min}$ &  &
    $-0.824_{-0.035}^{+0.042}$ & \\
   \hline
  $\kappa_\phi(\eta^\prime)_{\rm center}$ &
   $-0.844_{-0.016}^{+0.028}$ &
   $-0.872_{-0.033}^{+0.038}$ &
   $-0.893_{-0.038}^{+0.045}$ \\
   \hline
  $\kappa_\phi(\eta^\prime)_{\rm max}$ &  &
   $-0.915_{-0.032}^{+0.036}$ & \\
 \end{tabular}
\end{table}
%
%
\begin{figure}[hbp]
  \caption{ Leading diagrams
            in the chiral perturbation
            and $1/N_c$ expansion which contribute to the
            hadronic light-by-light scattering effect
            on the muon anomaly:
            (a)  pion-loop diagram,
            (b)  pseudoscalar pole diagram,
            (c)  quark-loop diagram.
            Solid and wavy lines represent
            muon and photon respectively.
            Dotted line in (a) corresponds
            to the charged pseudoscalar meson while
            dotted line connecting the two blobs in (b)
            corresponds to the neutral one.
            The closed solid line in (c)
            represents the quark loop.
            These diagrams
            are typical ones in respective classes. }
  \label{fig:diagram}
\end{figure}
\begin{figure}[hbp]
  \caption{  Pseudoscalar pole diagrams.
           Dotted line between the two blobs represents
           the propagation of neutral pseudoscalar meson. }
  \label{fig:pseudoscalarpole}
\end{figure}
\begin{figure}[hbp]
 \caption{ Comparison of various theoretical form factors of
          $ \Gamma(\pi^0 \rightarrow \gamma\gamma^*) $
          with the experimental data.
          Solid line corresponds to
          the VMD model ($b_1$)
          while dashed line ($b_2$) and dot-dashed lines ($b_3$)
          correspond to the QTL model and
          the QTL model supplemented by the VMD model,
          respectively. }
 \label{fig:formfactor}
\end{figure}
\begin{figure}[hbp]
 \caption{ Comparison of the various theoretical asymptotic
          behavior of form factors of
          $ \Gamma(\pi^0 \rightarrow \gamma\gamma^*) $ 
          with the experimental data.
          Three straight lines, from the above, are
          the Brodsky-Lepage (BL), OPE
          and the Gerard-Lahna (GL) predictions, respectively.
          The curve representing
          the Brodsky-Lepage interpolation
          formula ({\protect \ref{BLinterpolation}})
          is also shown for comparison. }
 \label{fig:asy}
\end{figure}
\vfill
\newpage
\pagestyle{empty}
\begin{figure}[hbp]
 \epsfbox{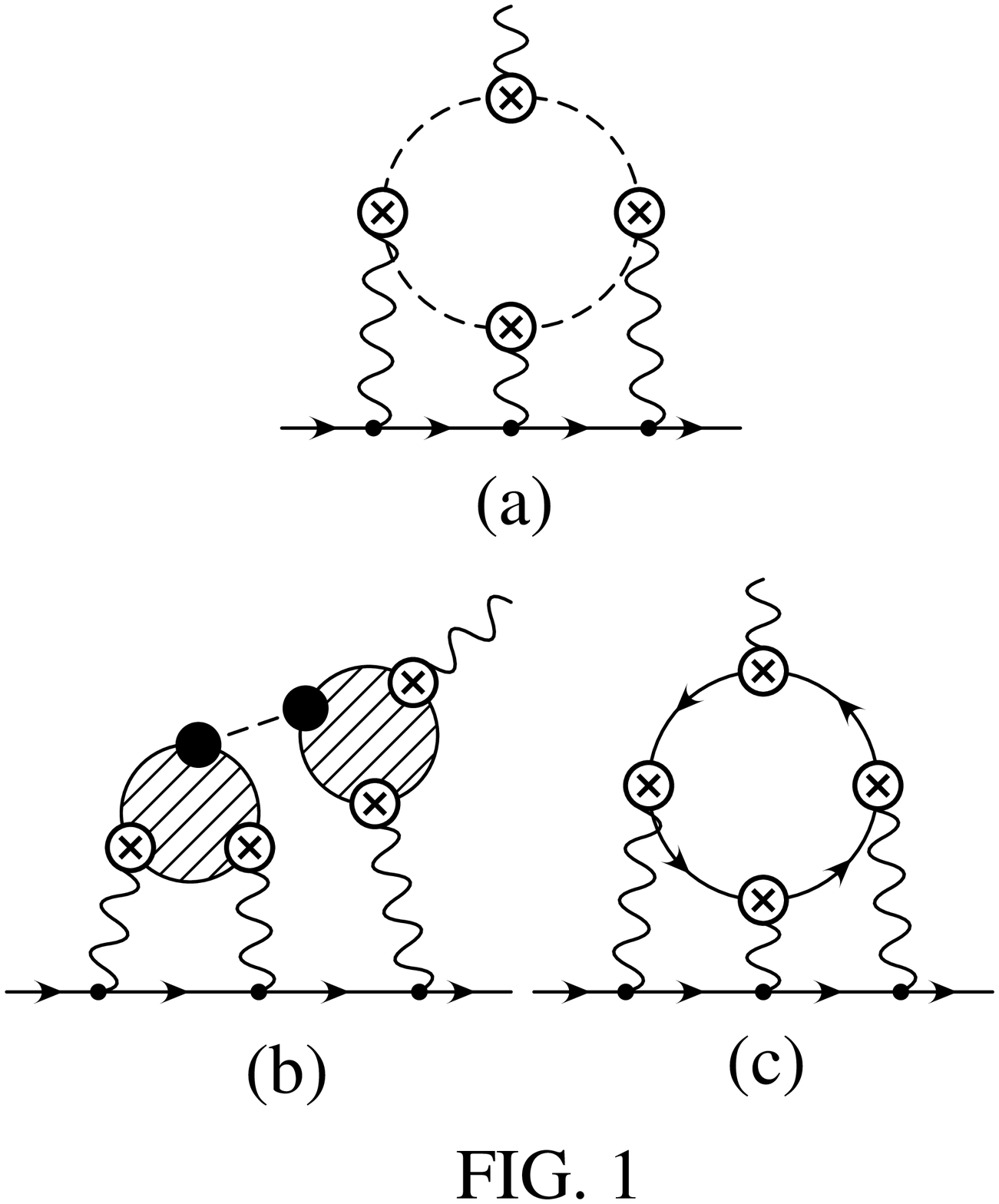}
\end{figure}
\vfill
\newpage
\begin{figure}[hbp]
 \epsfbox{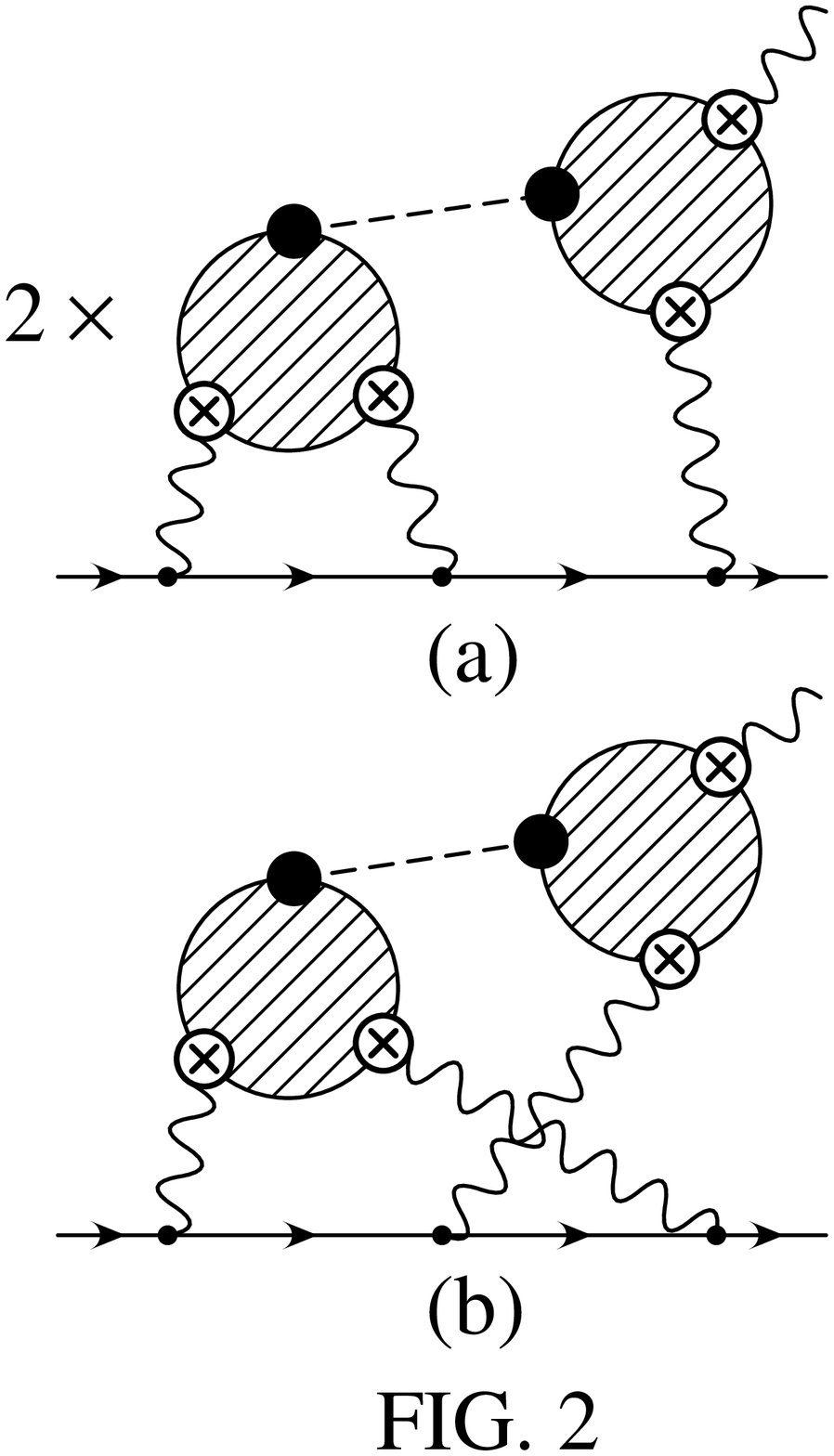}
\end{figure}
\vfill
\newpage
\begin{figure}[hbp]
 \epsfbox{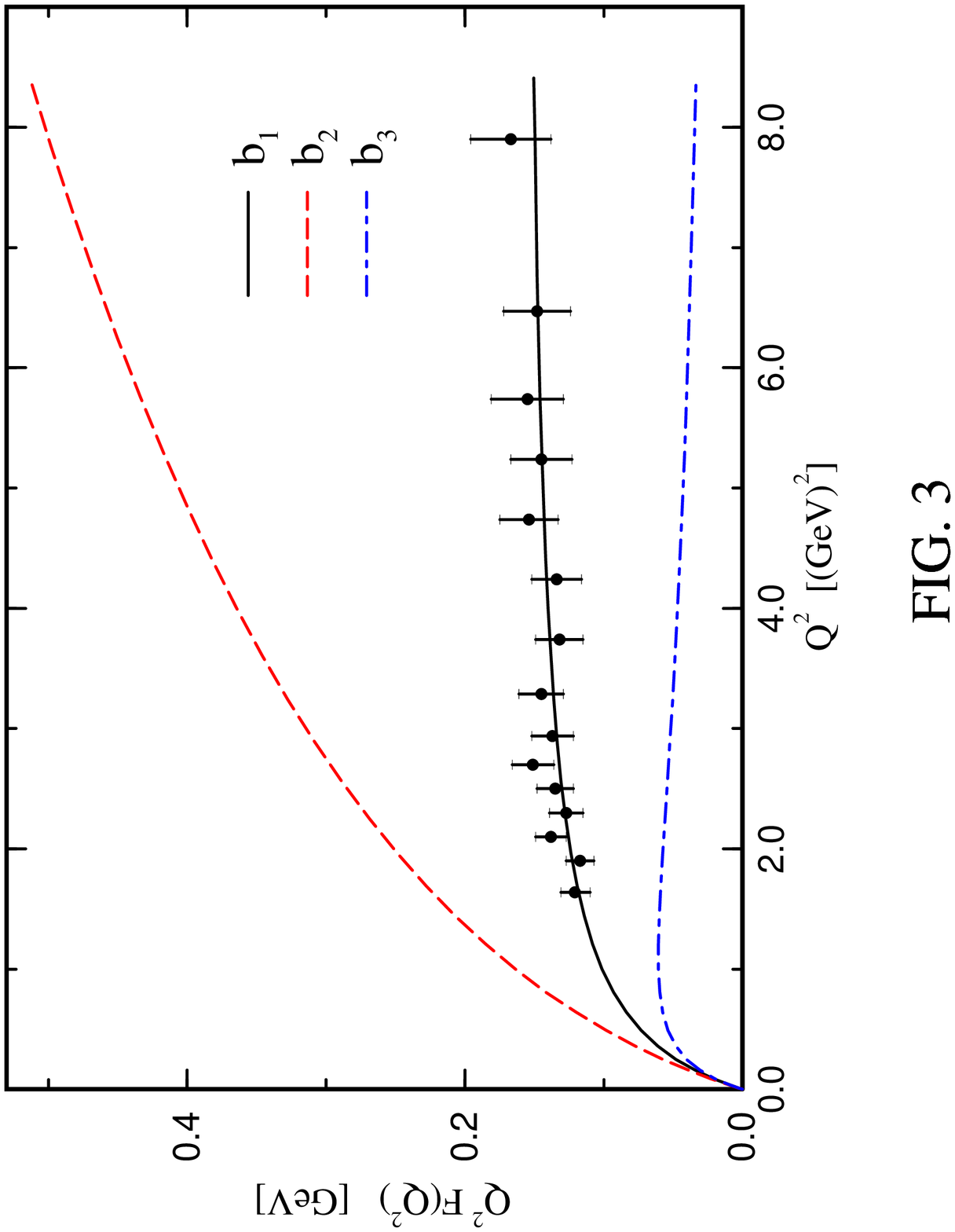}
\end{figure}
%
\newpage
\begin{figure}[hbp]
 \epsfbox{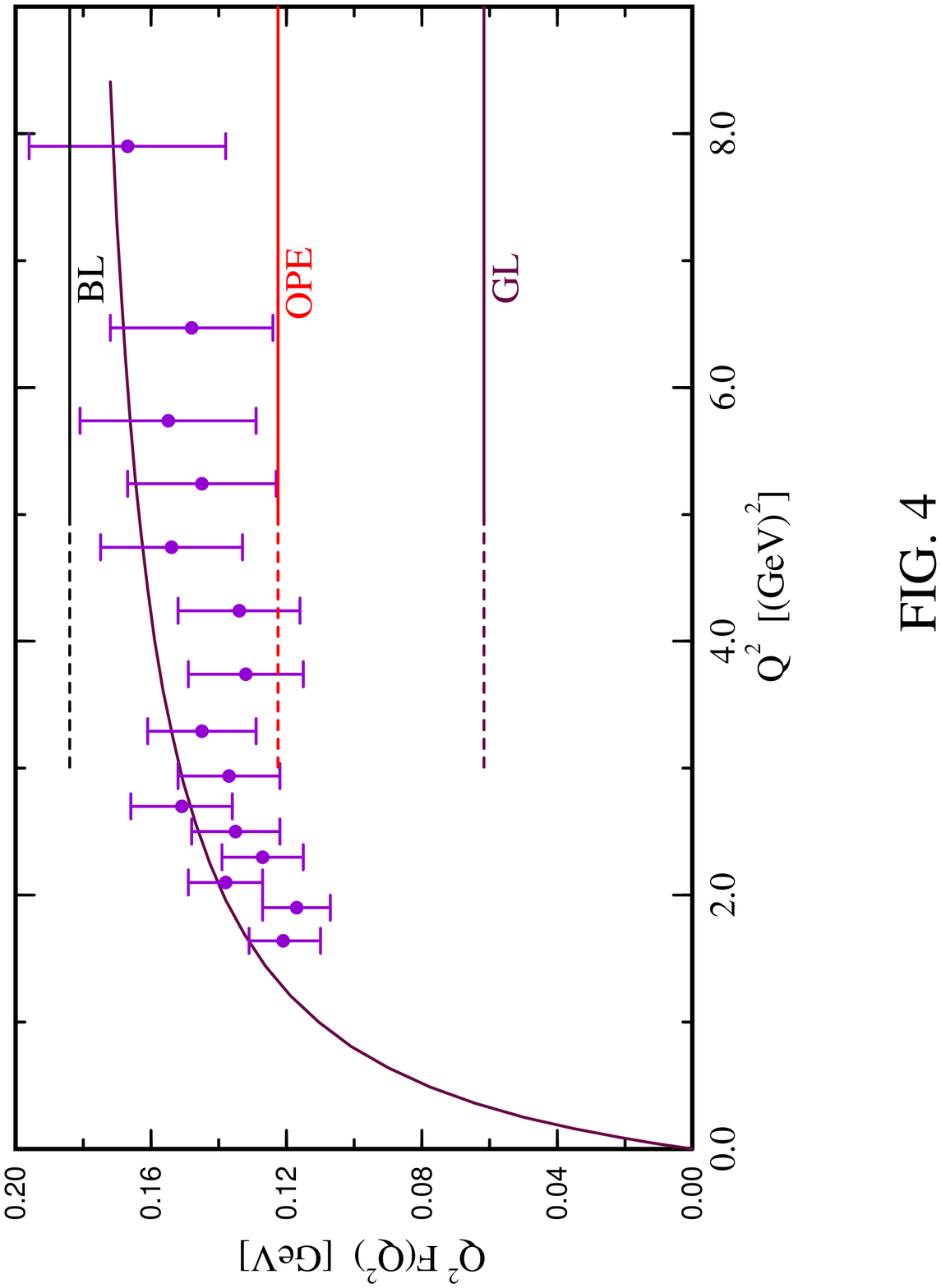}
\end{figure}
\end{document}